\documentclass[twocolumn,amsmath,amssymb,10pt,aps,superscriptaddress,nofootinbib,longbibliography]{revtex4} 
\usepackage[colorlinks=true]{hyperref}
\usepackage{graphicx}
\usepackage{soul}
\usepackage{latexsym}
\usepackage{amsmath}
\usepackage{amsthm}
\usepackage{amssymb}
\usepackage{bm}
\usepackage{xcolor}
\usepackage{subfigure}
\usepackage{pdfpages}
\usepackage{comment}
\usepackage{cleveref}
     \crefrangelabelformat{equation}%
       {(#3#1#4--#5\crefstripprefix{#1}{#2}#6)}
     \crefrangelabelformat{subequation}%
      {(#3#1#4--#5\crefstripprefix{#1}{#2}#6)}
\crefname{equation}{}{}

\allowdisplaybreaks 
\begin{document}
\title{Emergent Anisotropic Non-Fermi Liquid \\ at a Topological Phase Transition in Three Dimensions}

\author{SangEun Han}
\thanks{These authors contributed equally to this work.}
\affiliation{Department of Physics, Korea Advanced Institute of Science and Technology, Daejeon 305-701, Korea}
\author{Changhee Lee }
\thanks{These authors contributed equally to this work.}
\affiliation{Department of Physics and Astronomy, Seoul National University, Seoul 08826, Korea}
\author{Eun-Gook Moon}
\thanks{egmoon@kaist.ac.kr}
\affiliation{Department of Physics, Korea Advanced Institute of Science and Technology, Daejeon 305-701, Korea}
\author{Hongki Min}
\thanks{hmin@snu.ac.kr}
\affiliation{Department of Physics and Astronomy, Seoul National University, Seoul 08826, Korea}

\begin{abstract}
Understanding correlation effects in topological phases and their transitions is a cutting-edge area of research in recent condensed matter physics. 
We study topological quantum phase transitions (TQPTs) between double-Weyl semimetals (DWSMs) and insulators, and argue that a novel class of quantum criticality appears at the TQPT characterized by emergent {\it{anisotropic}} non-Fermi liquid behaviors, in which the interplay between the Coulomb interaction and electronic critical modes induces not only anisotropic renormalization of the Coulomb interaction but also strongly correlated electronic excitation in three spatial dimensions.
Using the standard renormalization group methods, large $N_f$ theory and the $\epsilon= 4-d$ method with fermion flavor number $N_f$ and spatial dimension $d$, 
we obtain the anomalous dimensions of electrons ($\eta_f=0.366/N_f $) in large $N_f$ theory and the associated anisotropic scaling relations of various physical observables. 
Our results may be observed in candidate materials for DWSMs such as HgCr$_2$Se$_4$ or SrSi$_2$ when the system undergoes a TQPT.
\end{abstract}

\date{\today}
\maketitle

{\em Introduction.} ---
Quantum criticality and topology play key roles in modern condensed matter physics  \cite{RevModPhys.82.3045,RevModPhys.83.1057,doi:10.1063/1.3554314,RevModPhys.90.015001,0034-4885-81-6-064501}, and the two concepts become naturally important near TQPTs. 
Recently, there has been a surge of interest in TQPTs  \cite{Bernevig1757,Konig766,Xu560,PhysRevB.85.195320,Wu2013,Faure2018}. The simplest class is described by the weakly interacting Dirac fermions, and it is well understood that the sign of the Dirac mass terms determines adjacent topological phases \cite{PhysRevLett.61.2015,PhysRevLett.95.226801,PhysRevLett.98.106803}. Since quasiparticles are well defined, non-interacting tight-binding models are sufficient to describe TQPTs in this class.

Beyond the simplest class, however, our understanding of TQPTs is far from complete. The long-range Coulomb interaction may drastically change the properties of non-interacting fermions near TQPTs, and the non-interacting tight-binding models {\it cannot} describe some classes of TQPTs. The interplay between critical electronic modes and the Coulomb interaction becomes significant, and quantum critical non-Fermi-liquid states may appear with emergent particle-hole and rotational symmetries \cite{PhysRev.102.1030,PhysRevB.69.235206,PhysRevLett.111.206401,PhysRevD.96.096010,PhysRevB.72.104404}. 
Moreover, the interplay may also give rise to weakly coupled but infinitely anisotropic excitations in a class of TQPTs \cite{PhysRevB.78.064512,PhysRevX.4.041027,Yang2014,Cho2016,PhysRevB.98.085149}. Thus, it is vital to deepen our understanding of TQPTs beyond the simplest class.

In this work, we uncover a novel class of TQPTs which shows emergent {\it{anisotropic}} non-Fermi-liquid behaviors in three spatial dimensions (3d) associated with topological nature of electronic wave functions.  
Our target system is the DWSM adjacent to insulator phases under the long-range Coulomb interaction. 
The presence of either the fourfold ($C_4$) or sixfold ($C_6$) rotational symmetry allows a direct phase transition between DWSMs and insulators whose bare Hamiltonian has a quadratic band touching spectrum. 
Without the symmetries, double-Weyl nodes may split into two Weyl nodes. The long-range Coulomb interaction becomes relevant at the critical point, and thus quasi-particle excitations are expected to be absent. Moreover, the absence of the cubic symmetry indicates a possibility of anisotropic quantum critical behaviors in contrast to most of the fixed points with the full rotational symmetry as in conformal field theories.
Using the standard renormalization group (RG) methods, we indeed find novel quantum critical phenomena with emergent {\it{anisotropy}}. For example, we find that the power-law dependences of the energy dispersion and the Coulomb interaction on momentum become anisotropic, even though they are initially set to be the same in all directions, and all excitations have anomalous dimensions.  
Our universality class is one concrete example of strongly interacting fixed points with non-Fermi-liquid behaviors beyond the conformal field theory description in 3d. We calculate its experimental signatures in physical observables such as the specific heat, compressibility, diamagnetic susceptibility, and optical conductivity.

\begin{figure}[b]
\centering
\includegraphics[width=0.8\linewidth]{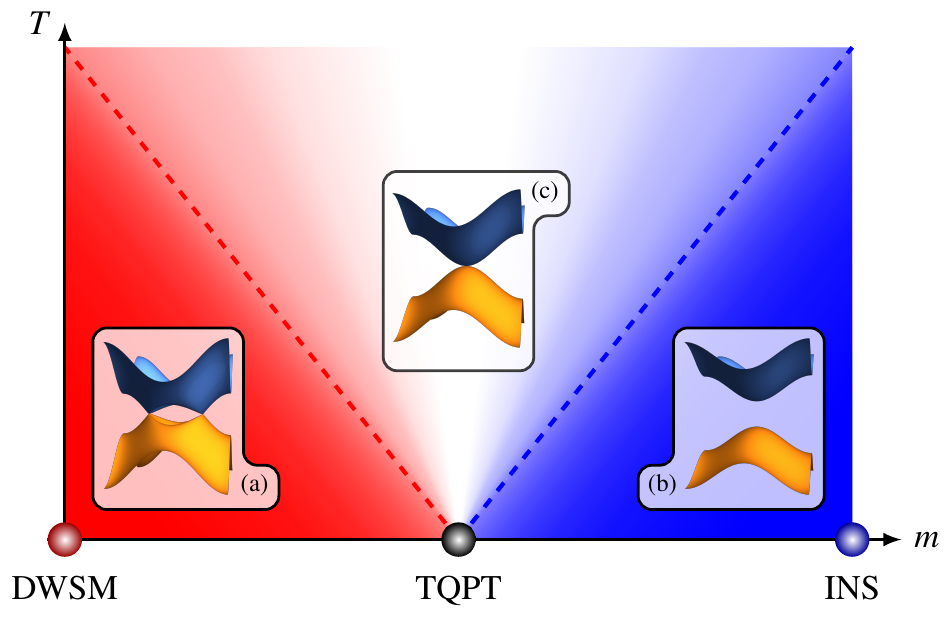}
\caption{Phase diagram for the TQPT between the DWSM and insulator phases with the tuning parameter $m$. 
The insets show the energy dispersions for the (a) DWSM, (b) insulator and (c) TQPT.
}\label{fig:phasediagram}
\end{figure}

{\em Model.} ---
We consider a minimal lattice model of DWSMs with a $C_4$ symmetry with a rotational axis along the {\it{z}} direction \cite{PhysRevB.95.201102,PhysRevLett.108.266802,PhysRevB.95.161112,PhysRevB.92.045121,PhysRevB.92.045121,SM},
\begin{align}
\label{eq:Lattice Hamiltonian}
\begin{split}
\mathcal{H}(\bm{k})=&2t'_{x}\left[{\cos(k_{y}a_0)-\cos(k_{x}a_0)}\right]\sigma_{x}\\
&+2t'_{y}\sin(k_{x}a_0)\sin(k_{y}a_0)\sigma_{y}+2M_{z}(\bm{k})\sigma_{z},
\end{split}
\end{align}
where $M_{z}(\bm{k})=m_{z}-t'_{z}\cos(k_{z}a_0)+m_{0}[2-\cos(k_{x}a_0)-\cos(k_{y}a_0)]$ and $a_0$ is the lattice constant. In general, $C_4$ symmetry does not imply $t'_{x}=t'_{y}$. However, in the presence of the Coulomb interaction, $t'_{x}=t'_{y}$ emerges at low energies \cite{SM}. For $|m_z|<t'_z$, the Hamiltonian supports two double-Weyl nodes at $\bm{k}=(0,0,\pm k_{z}^{*})$, where $k_{z}^{*}=a_{0}^{-1} \cos^{-1}(m_{z}/t'_{z})$, which are characterized by the Chern numbers $\pm2$ around the points \cite{PhysRevB.95.201102}. 
For $|m_{z}|>t'_{z}$, the system shows an insulator phase. At $|m_z|=t_z'$, a quantum phase transition occurs between the DWSM and insulator phases, as shown in Fig.~\ref{fig:phasediagram}. Neglecting $m_0$ for simplicity, we obtain the low-energy effective Hamiltonian near the transition point given by
\begin{align}
\mathcal{H}_0 (\bm{k})=&t_{\perp}[(k_{x}^{2}-k_{y}^{2})\sigma_{x}+2 k_{x}k_{y}\sigma_{y}]
+(t_{z}k_{z}^{2}+m)\sigma_{z},\label{eq:H_TQPT}
\end{align}
where $t_{\perp}=t'_{x}a_{0}^{2}$ and $t_{z}=t'_{z}a_{0}^{2}$. Here, a tuning parameter of the TQPT, $m\propto |m_{z}|-t'_{z}$, is introduced. The energy eigenvalues of the Hamiltonian are given by $E_{\pm}(\bm{k})=\pm \sqrt{t_{\perp}^2 (k_{x}^2+k_{y}^2)^2 + (t_{z} k_{z}^2 +m)^2}$, and 
at $m=0$ the energy dispersion becomes quadratic in all three directions.

The corresponding effective action with the long-range Coulomb interaction is
\begin{align}
\mathcal{S}=&\int d\tau d^{3}x\;\psi^{\dagger}[\partial_{\tau}-ig\phi+\mathcal{H}_{0}(-i\nabla)]\psi \\
&+\int d\tau d^{3}x\;\frac{1}{2}\left[a\left\{(\partial_{x}\phi)^{2}+(\partial_{y}\phi)^{2}\right\}+\frac{1}{a}(\partial_{z}\phi)^{2}\right], \nonumber
\end{align}
where $g\equiv\frac{\sqrt{4\pi}e_{0}}{\sqrt{\varepsilon}}$ with $e_0$ and $\varepsilon$ being the bare charge and the dielectric constant, respectively, $\psi$ is a spinor with $2N_{f}$ components, and $\phi$ is a bosonic field describing the long-range Coulomb interaction. 
Note that the topological aspect of Eq.~\eqref{eq:H_TQPT} ensures the band crossing associated with the $C_4$ symmetry of the system.
The parameter $a$ is introduced to characterize the anisotropy ratio of the Coulomb interaction between the $xy$-plane and the $z$-axis. For later usage, we define the following dimensionless parameters,
\begin{equation}
\alpha = \frac{A_{d-2}g^{2}}{\sqrt{t_{\perp}t_{z}}\Lambda^{4-d}},
\quad\beta=\frac{t_z}{t_{\perp}},
\quad\gamma=\frac{a\sqrt{\beta}}{2}\quad
\end{equation}
with $A_{d}=[6\pi(4\pi)^{\frac{d}{2}}\Gamma(\frac{d}{2})]^{-1}$. Here, $\alpha$ represents the ratio of the Coulomb potential and the electron kinetic energy, $\beta^{-1}$ is the anisotropy parameter for the fermionic fields, and $\gamma$ is the combination of the two anisotropy parameters $a$ and $\beta$.  We assume that all the four-fermion interactions, $u_{ijkl}\psi_{i}^{\dagger}\psi_{j}^{\dagger}\psi_{k}\psi_{l}$, are set to be small at the lattice-spacing scale and flow into the trivial fixed point as in the literature \cite{PhysRevB.75.235423}, which is also justified below and in the Supplemental Material \cite{SM}.

The bare scaling dimensions of the parameters can be obtained by setting $[\tau]=-z$, $[x,y]=-z_{\perp}$ and $[z]=-1$. We find $[\psi]=(2z_{\perp}+d-2)/2$, $[\phi]=(z+z_{\perp}+d-3)/2$, and $[g^{2}]=z-z_{\perp}-d+3$. Here, the dimension of the space along the {\it{z}}-axis is extended from the physical dimension of 1 to $d-2$. This extension is needed for the $\epsilon=4-d$ expansion method, as will be explained later. If we set ($t_z, t_{\perp}, a$) marginal, $z=2$ and $z_\perp=1$ thus, for $d=3$ the non-interacting fixed point becomes unstable because of the relevant coupling constant $g^2$, similar to that of the Luttinger-Abrikosov-Beneslavskii (LAB) phase \cite{abrikosov1970possible,abrikosov1974calculation,PhysRevLett.111.206401}. We stress that at our fixed point, the system becomes highly {\it{anisotropic}} distinguishing the scaling along the $z$ and $xy$ directions, which is fundamentally different from the {\it{isotropic}} LAB phase even though the bare scaling gives the same scaling dimensions. In this sense, our fixed point has emergent anisotropy.   

To deal with the relevant parameter, we employ the large $N_f$ method and the $\epsilon=4-d$ expansion method, and find that both methods give consistent results for the emergent anisotropic non-Fermi-liquid behaviors.

{\em Large $N_{f}$ calculation.} ---
We first use the large $N_f$ method since it is naturally extended from the conventional random phase approximation \cite{PhysRevB.93.165109,PhysRevLett.116.076803,PhysRevB.78.064512,PhysRevB.80.165102,PhysRevX.4.041027}. 
The boson self-energy is
\begin{align}
\Pi(i\Omega,\bm{q})=&N_{f}g^{2}\!\!\int_{\omega,\bm{k}}\!\!\!\!\text{Tr}[G_{0}(i\omega+i\Omega,\bm{k}+\bm{q})G_{0}(i\omega,\bm{k})], 
\end{align}
with the fermion propagator $G_{0}(i\omega,\bm{k})$ $=(-i \omega + \mathcal{H}_0(\bm{k}))^{-1}$. Here, we use the notation $\int_{\omega,\bm{k}}=\int\frac{d\omega}{2\pi}\frac{dk_{x}dk_{y}}{(2\pi)^{2}}\int'\frac{dk_{z}}{2\pi}$, where $\int'\frac{dk_{z}}{2\pi}$ stands for an integration over $\mu<|k_{z}|<\Lambda$ with the infrared (IR) cutoff $\mu$ and the ultra-violet (UV) cutoff $\Lambda$.
A detailed exposition of the boson self-energy is presented in the Supplemental Material \cite{SM}. We propose the following ansatz for the boson self-energy at one-loop level:
\begin{align}
\begin{split}
\Pi(i\Omega,\bm{q})=&-\frac{N_{f}g^{2}|q_{\perp}|}{\sqrt{t_{\perp}t_{z}}}F\left(\sqrt{\tfrac{t_{\perp}}{|\Omega|}}|q_{\perp}|,\sqrt{\tfrac{t_{z}}{t_{\perp}}}\left| \tfrac{q_{z}}{q_{\perp}}\right|\right) ,
\end{split}\label{eq:ansatz}
\end{align}
where $q_{\perp}\equiv \sqrt{q_{x}^2+q_{y}^2}$ and
\begin{align}
\begin{split}
F(x,y)=&\sqrt{C_{\perp_{1}}^{2}+C_{z_{1}}^{2}y^2}\tanh\left(x\sqrt{C_{\perp_{2}}^2+C_{z_{2}}^{2}y^2}\right),
\end{split}
\end{align}
with $C_{\perp_{1}}=0.041$, $C_{\perp_{2}}=1.199$, $C_{z_{1}}=0.016$, and $C_{z_{2}}=1.267$. For the details, see the Supplemental Material \cite{SM}.

The one-loop boson self-energy modifies the Coulomb potential in momentum space as
\begin{align}
D(i\Omega,\bm{q})=&\frac{1}{D_{0}^{-1}(\bm{q})-\Pi(i\Omega,\bm{q})},
\end{align}
where $D_{0}(\bm{q})=\left(a q_{\perp}^{2}+\frac{1}{a} q_{z}^{2}\right)^{-1}$ 
is the bare boson propagator. In the static ($\Omega=0$) and long wave length ($q\rightarrow 0$) limit,
the self-energy dominates the bare propagator since it linearly depends on the momentum in this limit.
Thus, we take the boson self-energy as the main contribution to the renormalized Coulomb interaction, $D(i\Omega,\bm{q}) \simeq \frac{1}{-\Pi(i\Omega, \bm{q})}$. 
This indicates that the boson is strongly renormalized from the quadratic to a linear momentum dependence, exhibiting the anomalous dimension of order one at the TQPT.
This approximation has been well established in large $N_f$ analysis and is checked afterward. 

The fermion self-energy with the renormalized Coulomb interaction is 
\begin{align}
\Sigma(i\omega,\bm{k})=&(-ig)^{2}\int_{\Omega,\bm{q}}\!\!\! G_{0}(i\omega+i\Omega,\bm{k}+\bm{q}) D(i\Omega,\bm{q}),
\end{align} 
and the fermion part of the action is modified by the fermion self-energy as
\begin{align}
-i \omega +\mathcal{H}_0(\bm{k}) \rightarrow -i \omega +\mathcal{H}_0(\bm{k}) - \Sigma(i\omega, \bm{k}).
\end{align}
It is straightforward to show that the corrections from the self-energy are logarithmically divergent in both UV and IR cutoffs, respectively, and we find
\begin{align}
\label{eq:self_energy_correction}
\begin{split}
\Sigma(i\omega,\bm{k}) \approx& \frac{C_{\omega}}{N_{f}}(i\omega)\ell-\frac{C_{t_{z}}}{N_{f}} \ell(t_{z}k_{z}^{2})\sigma_{z} \\
&-\frac{C_{t_{\perp}}}{N_{f}}\ell\left[t_{\perp}(k_{x}^{2}-k_{y}^{2})\sigma_{x}+2t_{\perp}k_{x}k_{y}\sigma_{y}\right] ,
\end{split}
\end{align}
where $C_{\omega}=0.366$, $C_{t_{\perp}}=0.614$, $C_{t_{z}}=0.341$, and $\ell=\log{\frac{\Lambda}{\mu}}$ is the RG parameter. For the details, see the Supplemental Material \cite{SM}. 

We also evaluate the vertex correction at vanishing external momentum and frequency,

\begin{align}
\delta_{g}=&(-ig)^{2}\int_{\Omega,\bm{q}}G_{0}(i\Omega,\bm{q})^{2}\frac{1}{-\Pi(i\Omega,\bm{q})}= \frac{C_{g}}{N_f}\ell,
\end{align}
where $C_{g}=0.366$, which is {\it exactly} the same as $C_{\omega}$. 
This agreement is not a coincidence but instead a consequence of the Ward identity $\delta_{g}={\partial \Sigma / \partial (i\omega)}$.

Using the logarithmic dependence of the self-energy, one can find various anomalous dimensions. 
The scale invariance at the critical point forces renormalization of the fermion fields with the anomalous dimension $\eta_f = \frac{C_{\omega}}{N_f}$. The non-zero anomalous dimension clearly indicates non-Fermi-liquid behaviors of the fermionic excitations, which can be understood by the absence of the pole structure in the fermionic Green function. 

From Eq.~(\ref{eq:self_energy_correction}), the RG equations for $t_{\perp}$ and $t_z$ are given by
\begin{align}
\label{eq:rg_tr_tz}
\frac{1}{t_{\perp}}\frac{d t_{\perp}}{d\ell}=\frac{C_{t_{\perp}}-C_\omega}{N_f},\quad\frac{1}{t_z}\frac{d t_z}{d\ell}=\frac{C_{t_{z}}-C_\omega}{N_f}.
\end{align}
From Eq.~(\ref{eq:rg_tr_tz}), we find $\frac{1}{\beta^{-1}}\frac{d\beta^{-1}}{d\ell}=\frac{C_{t_{\perp}}-C_{t_{z}}}{N_f}>0$, indicating that $\beta^{-1}$ diverges at the TQPT and that the fermionic excitations become highly anisotropic at low energies. Thus, our critical theory is described by an emergent anisotropic non-Fermi liquid.

\begin{figure}[t]
\centering
\subfigure[]{
\includegraphics[scale=0.8]{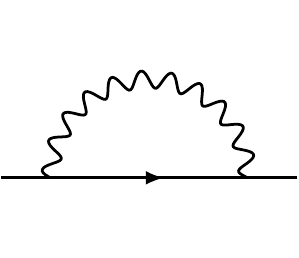}
\label{fig:1a}}
\subfigure[]{
\includegraphics[scale=0.8]{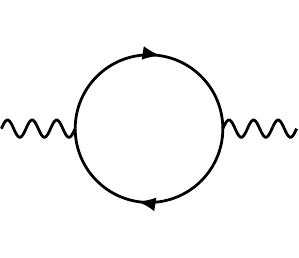}
\label{fig:1b}}
\subfigure[]{
\includegraphics[scale=0.8]{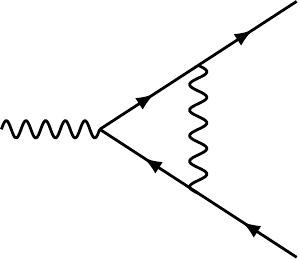}
\label{fig:1c}}
\caption{Feynman diagrams at one-loop order for the (a) fermion self-energy, (b) boson self-energy, and (c) vertex correction.
A straight line with an arrowhead and a wavy line represent the fermion and boson propagators, respectively. 
}\label{fig:loopdiagram}
\end{figure}

{\em{$\epsilon=4-d$ expansion.}}\label{Sec:epsilon expansion} --- Our large $N_f$ calculation is further supported by the standard $\epsilon=4-d$ expansion \cite{sachdev2011,PhysRevLett.111.206401,PhysRevB.94.195135,Lee2018}. Here, we introduce a new renormalization scheme in which the three spatial dimensions are embedded into a manifold that has more coordinates in the direction of the rotational axis ($z$-direction). Namely, we extend the coordinates as
\begin{align} 
\int \frac{dk_{x}dk_{y}}{(2\pi)^{2}} \int\frac{dk_{z}}{2\pi} 
\rightarrow \int\frac{dk_{x}dk_{y}}{(2\pi)^{2}}\int\frac{dk_{z}d^{d-3}p}{(2\pi)^{d-2}}
\end{align}
with $k_{z}^{2}\rightarrow k_{z}^{2}+p^{2}$, and the momentum $\bm{p}$ lives in a $(d-3)$-dimensional manifold. 
Recalling $[g^2]=z-z_{\perp}+3-d$ with $z=2$ and $z_\perp=1$, the coupling constant 
becomes marginal at $d=4$ and the quantum fluctuations give logarithmic divergences. To read off these logarithmic divergences, we introduce the parameter $\epsilon=4-d$ and employ the standard momentum shell RG analysis with $\epsilon$ expansion.
For the momentum shell integration, we impose the UV and IR cutoffs on the $(d-2)$-dimensional space of $(k_z,\bm{p})$ as
\begin{align}
\begin{split}
\int_{\bm{k},{\bm{p}}}=&\int{\frac{dk_xdk_y}{(2\pi)^2}}\int_{\partial \Lambda}{\frac{dk_zd^{d-3}p}{(2\pi)^{d-2}}},
\end{split}
\end{align}
where $\partial \Lambda$ represents an infinitesimal momentum shell $\mu<\sqrt{k_{z}^2+p^2}<\Lambda$ with $\mu=\Lambda e^{-\ell}$.
 
By integrating out the high energy modes, we obtain corrections at one-loop order. The fermion self-energy depicted by the diagram in Fig.~\ref{fig:1a} is given by
\begin{align}
\Sigma(i\Omega,\bm{q})=&(-ig)^{2}\int_{\omega,\bm{k},\bm{p}}\!\! G_{0}(i\omega+i\Omega,\bm{k}+\bm{q})D_{0}(i\omega,\bm{k}) \nonumber\\
\approx&-\alpha F_{\perp}(\gamma)\ell\left[t_{\perp}(q_{x}^{2}-q_{y}^{2})\sigma_{x}+2t_{\perp}q_{x}q_{y}\sigma_{y} \right] \nonumber \\
&-\alpha F_{z}(\gamma)\ell\left(t_{z}q_{z}^{2}\right)\sigma_{z},
\end{align}
where $F_\perp$ and $F_z$ are dimensionless functions, whose explicit expressions are presented in the Supplemental Material \cite{SM}.
Note that the frequency part is not renormalized at the one-loop order because of the instantaneous nature of the bare Coulomb interaction. Then, it is easy to see that the vertex correction [Fig.~\ref{fig:1c}] vanishes due to the Ward identity.
For the boson self-energy [Fig.~\ref{fig:1b}], we find
\begin{align}
\Pi(\bm{q})=&N_{f}g^{2}\int_{\omega,\bm{k},\bm{p}}\!\! \text{Tr}\left[G_{0}(i\omega,\bm{k}+\bm{q}/2)G_{0}(i\omega,\bm{k}-\bm{q}/2)\right] \nonumber \\
\approx&-N_{f}\alpha\left[	\frac{a}{\gamma} q_{\perp}^2+\frac{\gamma}{a}q_{z}^{2}	\right]\ell.
\end{align}

\begin{figure}[t]
\centering
\subfigure{
\includegraphics[scale=1]{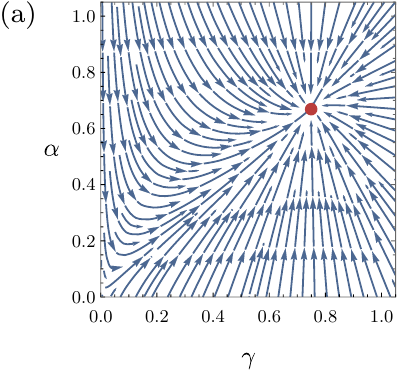}
\label{fig:rgflow01}}\hspace{-0.25em}
\subfigure{
\includegraphics[scale=1]{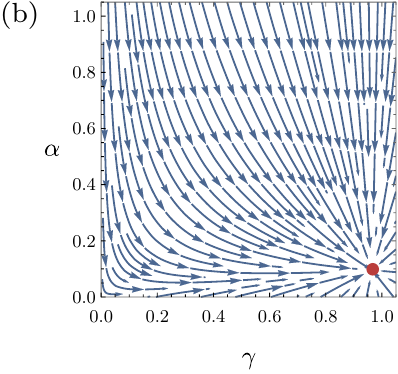}
\label{fig:rgflow10}}
\caption{RG flows of $\alpha$ and $\gamma$ for (a) $N_{f}=1$ and (b) $N_{f}=10$ at $\epsilon=1$. The red dots represent the fixed points $(\alpha^{*},\gamma^{*})$. For $N_{f}=1$, $(\alpha^{*},\gamma^{*})=(0.671,0.748)$ and for $N_{f}=10$, $(\alpha^{*},\gamma^{*})=( 0.096,0.966)$ obtained from Eq.~(\ref{eq:RG_alpha_beta}).
}\label{fig:rgflow}
\end{figure}

Renormalizing the wave functions and the coupling constants, we obtain the RG equations for $\alpha$ and $\gamma$ as
\begin{align}
\begin{split}
\label{eq:RG_alpha_beta}
\frac{1}{\alpha}\frac{d\alpha}{d\ell}&=\epsilon-\frac{N_{f}\alpha}{2}\left(	\frac{1}{\gamma}+\gamma	\right)-\frac{\alpha}{2}F_{+}(\gamma),\\
\frac{1}{\gamma}\frac{d\gamma}{d\ell}&=\frac{N_{f}\alpha}{2}\left(	\frac{1}{\gamma}-\gamma\right)+\frac{\alpha}{2}F_{-}(\gamma),
\end{split}
\end{align}
where $F_\pm(x)=F_{z}(x)\pm F_{\perp}(x)$. We find two fixed points from the RG equations in Eq.~(\ref{eq:RG_alpha_beta}). The non-interacting fixed point $\alpha^{*}=0$ with arbitrary $\gamma^{*}$ is unstable, whereas there exists a stable interacting fixed point at $(\alpha^*$,$\gamma^*)$ with $\alpha^{*}>0$. For $N_{f}=1$ and $\epsilon=1$, the stable fixed point is located at $(\alpha^{*},\gamma^{*})=(0.671,0.748)$, and for large $N_f$, $(\alpha^{*}$,$\gamma^{*})\approx(\epsilon/N_{f},1-0.358/N_{f})$. The RG flows of $\alpha$ and $\gamma$ are illustrated in Fig.~\ref{fig:rgflow}.

At the stable fixed point, the RG equations for the bosonic and fermionic anisotropy parameters are given by, respectively,
\begin{align}
\begin{split}
\left.\frac{1}{a}\frac{da}{d\ell}\right|_{\text{f.p.}}=&\frac{N_{f}\alpha^*}{2}\left(	\frac{1}{\gamma^*}-\gamma^*	\right) > 0, \\
\left.\frac{1}{\beta^{-1}}\frac{d\beta^{-1}}{d\ell}\right|_{\text{f.p.}}=&-\alpha^{*}F_{-}(\gamma^*) > 0,
\end{split}
\end{align}
where f.p. stands for the fixed point. Note that $\beta^{-1}$ diverges at the stable fixed point as in the large $N_f$ calculation demonstrating an emergent anisotropic non-Fermi liquid, which becomes a sanity check of our analysis giving a consistent result with the large $N_f$ calculation. For the details, see the Supplemental Material \cite{SM}.

{\em Physical observables.} ---
Recently, several materials \cite{PhysRevLett.108.266802,PhysRevLett.107.186806,Huang1180,PhysRevB.96.041206,Singh2018} have been proposed as possible candidates for DWSMs, in which TQPTs may occur by tuning the system parameters. For example, it has been theoretically demonstrated that SrSi$_2$ can be tuned by changing the lattice constant through doping or strain, leading to a transition from the DWSM to a trivial insulator phase \cite{Singh2018}. Since the anisotropic non-Fermi-liquid behavior at the TQPT will provide power-law corrections anisotropically to the scaling of physical observables \cite{PhysRevLett.116.076803,PhysRevLett.99.226803,PhysRevLett.108.046602}, the anisotropic scaling relations will be valuable to experiments.

First, consider the parameter dependence of physical observables in the non-interacting limit \cite{PhysRevB.95.161112,SM,2018arXiv180209050W,2018arXiv180210365W,PhysRevB.91.235131,PhysRevB.92.045121,FUKUYAMA1970111}. The details are presented in the Supplemental Material \cite{SM}. In the non-interacting limit, the specific heat $C_V$, compressibility $\kappa$, diamagnetic susceptibility $\chi_D$, and optical conductivity $\sigma$ are give by
\begin{align}
&&C_{V}\propto\;&\frac{T^{3/2}}{t_{\perp}t_{z}^{1/2}},&\quad\kappa\propto\;& \frac{T^{1/2}}{t_{\perp}t_{z}^{1/2}},&&\\
&&\chi_{D,\perp}\propto\;& t_{z}^{1/2}T^{1/2},&\quad\chi_{D,z}\propto\;& \frac{t_{\perp}}{t_{z}^{1/2}}T^{1/2}, &&\nonumber \\
&&\sigma_{\perp\perp}\propto\;& \frac{1}{t_{z}^{1/2}}\Omega^{1/2},&\quad\sigma_{zz}\propto\;& \frac{t_{z}^{1/2}}{t_{\perp}}\Omega^{1/2}. &&\nonumber
\end{align}
Here, $\chi_{D,x}=\chi_{D,y}=\chi_{D,\perp}$ and $\sigma_{xx}=\sigma_{yy}=\sigma_{\perp\perp}$ because of the $C_4$ symmetry of the Hamiltonian. We also assume $t_{x} =t_{y}= t_{\perp}$ for simplicity.

Now, consider how the anisotropic non-Fermi liquids change the bare scaling behaviors of the physical observables.
From the $\epsilon$ expansion, the RG equations for $t_{\perp}$ and $t_{z}$ are given by
\begin{align}
\label{eq:beta_tr_tz}
\begin{split}
\frac{1}{t_{\perp}}\frac{d t_{\perp}}{d\ln b}=&z-2z_{\perp}+\alpha F_{\perp}(\gamma), \\
\frac{1}{t_z}\frac{d t_{z}}{d\ln b}=&z-2+\alpha F_{z}(\gamma),
\end{split}
\end{align}
where $\ln b\equiv\ell$. 
Let us choose $z=2$ and $z_{\perp}=1$ so that $t_{\perp}$ and $t_z$ are marginal at the tree level. Since $\frac{d\mathcal{O}}{d\ln b}=z\mathcal{O}$ for $\mathcal{O}=\omega,T$ with $z=2$, $\mathcal{O}(b)=\mathcal{O}b^{2}$. Let $b^{*}$ be the cutoff value defined as $\mathcal{O}(b^{*})=\Lambda$, then $b^{*}=(\Lambda/\mathcal{O})^{1/2}$. Combining this with Eq.~(\ref{eq:beta_tr_tz}), we find that $t_{i}(b^{*})=t_{i,0}(b^{*})^{\alpha^{*}F_{i}(\gamma^{*})}\propto \mathcal{O}^{-c_{i}}$ where $i=\perp,z$, $c_{i}=\frac{1}{2}\left.\frac{d\ln t_{i}}{d\ln b}\right|_{\text{f.p.}}=\alpha^{*}F_{i}(\gamma^{*})/2$, $c_{\perp}\approx0.402/N_{f}$, and $c_{z}\approx0.044/N_{f}$ in the large $N_{f}$ approximation.

Then, near the interacting fixed point, the scaling relations of the physical observables with respect to either temperature or frequency become
\begin{align}
&&C_{V}\propto\;& T^{3/2+\eta_{1}},\quad &\kappa\propto\;& T^{1/2+\eta_{1}},&&\\
&&\chi_{D,\perp}\propto\;& T^{1/2-\eta_{2}},\quad& \chi_{D,z}\propto\;& T^{1/2-\eta_{3}},&&\nonumber \\
&&\sigma_{\perp\perp}\propto\;&\Omega^{1/2+\eta_{2}},\quad& \sigma_{zz}\propto\;&\Omega^{1/2+\eta_{3}}, &&\nonumber
\end{align}
where $\eta_{1}\equiv c_{\perp}+c_{z}/2\approx0.423/N_{f}$, $\eta_{2}\equiv c_{z}/2\approx0.022/N_{f}$, and $\eta_{3}\equiv c_{\perp}-c_{z}/2\approx 0.380/N_{f}$.
(Equivalently, we can obtain the same results by including all the effects of renormalization in the coordinates rather than the system parameters, as presented in the Supplemental Material \cite{SM}.)
Thus, it is easily seen that the diamagnetic susceptibility and optical conductivity show anisotropic scaling behaviors, $\chi_{D,z}/\chi_{D,\perp}\propto T^{\eta_{2}-\eta_{3}}$ and $\sigma_{\perp\perp}/\sigma_{zz}\propto \Omega^{\eta_{2}-\eta_{3}}$.
In addition, the permittivity tensor characterizing the charge screening also exhibits the anisotropic behavior, $\varepsilon_{\perp}/\varepsilon_{z}=a^2 \propto \Omega^{\eta_{2}-\eta_{3}}$. By measuring these ratios, we can clearly see the anisotropic scaling behaviors at the TQPT.

{\em Discussion and Conclusion.} ---
So far, for simplicity we ignored $m_0$ in Eq.~(\ref{eq:Lattice Hamiltonian}) and the corresponding $s_{\perp}(k_{x}^2+k_{y}^2)\sigma_z$ term with $s_{\perp}=m_0 a_{0}^{2}$ in $\mathcal{H}_{0}$, which is allowed by symmetry. If we include the effect of this term, we find that there still exists a stable non-Gaussian fixed point at $(\alpha^{*},\gamma^{*},\lambda^{*})=(0.336\epsilon/N_{f}, 0.821-0.083/N_{f}, -\text{sgn}(\beta)(0.866+0.035/N_{f}))$ in the $\epsilon$ expansion ($\lambda\equiv s_{\perp}/t_{\perp}$), indicating that the anisotropic non-Fermi-liquid behavior is robust against the $s_{\perp}(k_{x}^2+k_{y}^2)\sigma_z$ term. The details are presented in the Supplemental Material \cite{SM}. Note that for a TQPT between triple-Weyl semimetals \cite{2018arXiv180210365W}, we believe that similar symmetry-allowed parabolic term should be considered.

There are studies on the instability of NFL in the quadratic dispersions under the presence of the short-range interactions \cite{PhysRevLett.113.106401,PhysRevB.92.045117}. 
Our calculations are controlled by either $\epsilon$ or $1/N_f$. Thus, the scaling dimensions of the four-point short-range interactions at the stable fixed point are the same as the bare one at the leading order, $[u_{ijkl}]=-d+2+\mathcal{O}\;(\epsilon\text{ or }1/N_{f})$, which indicates that our fixed point is stable under the short-range interactions.  A detailed discussion on the effect of the short-range interactions can be found in the Supplemental Material \cite{SM}. 

We stress that our emergent anisotropic non-Fermi-liquid fixed point is distinct from previously studied non-Fermi-liquid fixed points. 
Our fixed point is in 3d in sharp contrast to most of the previously studied fixed points including the very nice work by Sur and Lee where anisotropic non-Fermi liquid below 3d was found \cite{PhysRevB.94.195135}. 
In 3d, quantum fluctuations are typically marginal or even irrelevant, so quasi-particles are usually well-defined. However, the interplay between the topology and $C_4$ rotational symmetry in our systems protects the quadratic band touching at the topological phase transition, and the anisotropic non-Fermi-liquid fixed point appears. As discussed above, the absence of the cubic symmetry makes the anisotropy even emergent in terms of the anomalous dimensions. 
Furthermore, the characteristic interplay between topology and symmetry is crucial in addition to the long-range Coulomb interaction to realize our universality class.

In summary, we studied TQPTs between DWSMs and insulators using the large $N_f$ theory and epsilon expansion. We found that a novel class of quantum criticality appears at the TQPT characterized by emergent anisotropic non-Fermi-liquid behaviors in which critical electronic modes and the long-range Coulomb interaction are strongly coupled, and the system becomes infinitely anisotropic. The anisotropic behaviors at the TQPT may be observed experimentally by measuring the power-law corrections to the diamagnetic susceptibility $\chi_{D,z}/\chi_{D,\perp}\propto T^{\eta_{2}-\eta_{3}}$ and optical conductivity $\sigma_{\perp\perp}/\sigma_{zz}\propto\Omega^{\eta_{2}-\eta_{3}}$, which we propose as smoking-gun signals of our TQPTs.   

\begin{acknowledgments}
We thank S. X. Zhang, S. K. Jian, and H. Yao for sharing their preprint \cite{2018arXiv180910686Z}. 
C.L. and H.M. were supported by the National Research Foundation of Korea (NRF) grant funded by the Korea government (MSIT) (No. 2018R1A2B6007837) and Creative-Pioneering Researchers Program through Seoul National University (SNU). S.E. Han and E.-G. Moon were supported by the POSCO Science Fellowship of POSCO TJ Park Foundation and NRF of Korea under Grant No. 2017R1C1B2009176.
\end{acknowledgments}

%

\onecolumngrid
\clearpage
\begin{center}
\textbf{\large Supplemental Material for ``Emergent Anisotropic Non-Fermi Liquid \\
at a Topological Phase Transition in Three Dimensions''}
\end{center}
\begin{center}
{SangEun Han,$^{1,*}$ Changhee Lee,$^{2,*}$ Eun-Gook Moon,$^{1,\dagger}$ and Hongki Min$^{2,\ddagger}$}\\
\emph{$^{1}$Department of Physics, Korea Advanced Institute of Science and Technology, Daejeon 305-701, Korea}\\
\emph{$^{2}$Department of Physics and Astronomy, Seoul National University, Seoul 08826, Korea}
\end{center}
\setcounter{equation}{0}
\setcounter{figure}{0}
\setcounter{table}{0}
\setcounter{page}{1}

\makeatletter
\renewcommand{\thesection}{\Roman{section}}
\renewcommand{\thesubsection}{
\Alph{subsection}}
\renewcommand{\thesubsubsection}{
\arabic{subsubsection}}
\renewcommand{\theequation}{S\arabic{equation}}
\renewcommand{\thefigure}{S\arabic{figure}}

\section{Details of the $\epsilon=4-d$ method}
\label{app:c4}

In this section, we provide detailed calculations of the $\epsilon=4-d$ method. First, we prove that $t_x=t_y$ and $a_x=a_y$ at low energies. Next, we derive the renormalization group (RG) equations using the $\epsilon=4-d$ expansion. Then we discuss the effect of the symmetry-allowed parabolic term, which is neglected in the main text, demonstrating that the TQPT is still characterized by anisotropic non-Fermi liquids.

Consider the leading-order self-energy corrections for fermions and bosons:
\begin{align}
\Sigma(0,\bm{k})=&(-ig)^{2}\int_{\Omega,\bm{q}}G_{0}(i\Omega,\bm{q}+\bm{k})D_{0}(i\Omega,\bm{q}),\\
\Pi(i\Omega,\bm{q})=&-N_{f}(-ig)^{2}\int_{\omega,\bm{k}}\text{Tr}[G_{0}(i\Omega+i\omega,\bm{k}+\bm{q}/2)G_{0}(i\omega,\bm{k}-\bm{q}/2)],
\end{align}
where $\int_{\Omega,\bm{q},\bm{p}}=\int_{\Omega}\frac{d\Omega}{2\pi}\int\frac{dq_{x}dq_{y}}{(2\pi)}\int_{\partial\Lambda}\frac{dq_{z}d^{d-3}p}{(2\pi)^{d-2}}$ with $\partial \Lambda$ being the region $\mu<\sqrt{q_{z}^2+p^2}<\Lambda$. Here,
\begin{align}
G_{0}(i\Omega,\bm{k})=&\frac{1}{-i\Omega+\varepsilon_{x}(\bm{k})\sigma_{x}+\varepsilon_{y}(\bm{k})\sigma_{y}+\varepsilon_{z}(\bm{k})\sigma_{z}}=\frac{i\Omega+\varepsilon_{x}(\bm{k})\sigma_{x}+\varepsilon_{y}(\bm{k})\sigma_{y}+\varepsilon_{z}(\bm{k})\sigma_{z}}{\Omega^{2}+E(\bm{k})^2},\\
D_{0}(i\Omega,\bm{q})=&\frac{1}{a_{x}q_{x}^{2}+a_{y}q_{y}^{2}+a_{z}q_{z}^{2}},
\end{align}
where $\varepsilon_{x}(\bm{k})=t_{x}(k_{x}^{2}-k_{y}^{2})$, $\varepsilon_{y}(\bm{k})=2t_{y}k_{x}k_{y}$, $\varepsilon_{z}(\bm{k})=t_{z}k_{z}^{2}$, and $E(\bm{k})=\sqrt{\varepsilon_{x}(\bm{k})^{2}+\varepsilon_{y}(\bm{k})^{2}+\varepsilon_{z}(\bm{k})^{2}}$.

\subsection{Proof of the emergent rotational symmetry along the $k_{z}$-axis}
\subsubsection{Proof of $a_{x}=a_{y}$}\label{app:aa}
First, let us prove that $a_x=a_y$ at low energies. From the self-energy of the Coulomb interaction at $\Omega=0$, 
\begin{align}
\label{eq:Pi}\Pi(0,\bm{k})=&-N_{f}(-ig)^{2}\int_{\omega,\bm{q}}\text{Tr}[G_{0}(i\omega,\bm{q}+\bm{k}/2)G_{0}(i\omega,\bm{q}-\bm{k}/2)]\nonumber \\
=&-N_{f}g^{2}\int_{\bm{q},\bm{p}}\left(1-\frac{\vec{\varepsilon}_{+}\cdot\vec{\varepsilon}_{-}}{E_{+}E_{-}}\right)\frac{1}{E_{+}+E_{-}}\nonumber \\
\approx&-N_{f}g^{2}\int_{\bm{q},\bm{p}}\left[\frac{1}{a_{x}}\frac{(q_{x}^{2}+q_{y}^{2})(t_{x}^{2}t_{y}^{2}(q_{x}^{2}+q_{y}^{2})^{2}+t_{z}^{2}(t_{x}^{2}+t_{y}^{2})(q_{z}^{2}+p^{2})^{2})}{2(t_{x}^{2}(q_{x}^{2}-q_{y}^{2})^{2}+4t_{y}^{2}q_{x}^{2}q_{y}^{2}+t_{z}^{2}(q_{z}^{2}+p^{2})^{2})^{5/2}}a_{x}k_{x}^{2}\right.\nonumber\\
&\quad\quad\quad\quad\quad\quad\left.+\frac{1}{a_{y}}\frac{(q_{x}^{2}+q_{y}^{2})(t_{x}^{2}t_{y}^{2}(q_{x}^{2}+q_{y}^{2})^{2}+t_{z}^{2}(t_{x}^{2}+t_{y}^{2})(q_{z}^{2}+p^{2})^{2})}{2(t_{x}^{2}(q_{x}^{2}-q_{y}^{2})^{2}+4t_{y}^{2}q_{x}^{2}q_{y}^{2}+t_{z}^{2}(q_{z}^{2}+p^{2})^{2})^{5/2}}a_{y}k_{y}^{2}\right]\nonumber\\
&\quad\quad\quad\quad\quad\quad\left.+\frac{1}{a_{z}}\frac{t_{z}^{2}q_{z}^{2}(t_{x}^{2}(q_{x}^{2}-q_{y}^{2})^{2}+4t_{y}^{2}q_{x}^{2}q_{y}^{2})}{(t_{x}^{2}(q_{x}^{2}-q_{y}^{2})^{2}+4t_{y}^{2}q_{x}^{2}q_{y}^{2}+t_{z}^{2}(q_{z}^{2}+p^{2})^{2})^{5/2}}a_{z}k_{z}^{2}\right],
\end{align}
where $\varepsilon_{i\pm}=\varepsilon_{i}(\bm{q}\pm\bm{k}/2)$ and $E_{\pm}=\sqrt{\sum_{i}{{\varepsilon}_{i\pm}^{2}}}$.\\
We find that the coefficients of the $k_{x}^2$ and $k_{y}^2$ terms are the same, which we denote as $C_{a}$, are given by
\begin{align}
C_{a}=&-N_{f}g^{2}\int_{\bm{q},\bm{p}}\frac{(q_{x}^{2}+q_{y}^{2})\left[t_{x}^{2}t_{y}^{2}(q_{x}^{2}+q_{y}^{2})^{2}+t_{z}^{2}(t_{x}^{2}+t_{y}^{2})(q_{z}^{2}+p^{2})^{2}\right]}{2\left[t_{x}^{2}(q_{x}^{2}-q_{y}^{2})^{2}+4t_{y}^{2}q_{x}^{2}q_{y}^{2}+t_{z}^{2}(q_{z}^{2}+p^{2})^{2}\right]^{5/2}}\nonumber\\
\propto&-\frac{N_{f}g^{2}}{\Lambda^{4-d}}\ell,
\end{align}
where $\ell=\ln({\Lambda/\mu})$. Let $C'_{a}=-C_{a}/\ell$, which is positive regardless of $t_x$, $t_y$ and $t_z$. Then, the beta function of $a_{x}/a_{y}$ is
\begin{align}
\frac{1}{a_{x}/a_{y}}\frac{d(a_{x}/a_{y})}{d\ell}=&C'_{a}a_{y}\left(1-\frac{a_{x}}{a_{y}}\right).
\end{align}
Since $C'_{a}$ is positive, $a_{x}=a_{y}$ at low energies.

\subsubsection{Proof of $t_{x}=t_{y}$}\label{app:txtytr}
From now on, we employ the following form of the Coulomb interaction propagator with $a_x=a_y\equiv a$ and $a_z=1/a$,
\begin{align}
D_{0}(i\Omega,\bm{q})=&\frac{1}{a(q_{x}^{2}+q_{y}^{2})+(q_{z}^{2}+p^{2})/a}.
\end{align}
Then
\begin{align}
\Sigma(i\omega,\bm{k})=&(-ig)^{2}\int_{\Omega,\bm{q},\bm{p}}G_{0}(i\omega+i\Omega,\bm{k}+\bm{q})D_{0}(i\Omega,\bm{q}),\nonumber\\
=&-\frac{g^{2}}{2}\int_{\bm{q},\bm{p}}\frac{\varepsilon_{x}(\bm{k}+\bm{q})\sigma_{x}+\varepsilon_{y}(\bm{k}+\bm{q})\sigma_{y}+\varepsilon_{z}(\bm{k}+\bm{q})\sigma_{z}}{E(\bm{k}+\bm{q})}\frac{1}{a(q_{x}^{2}+q_{y}^{2})+(q_{z}^{2}+p^{2})/a},\nonumber\\
\approx&-\delta_{t_{x}}\varepsilon_{x}(\bm{k})\sigma_{x}-\delta_{t_{y}}\varepsilon_{y}(\bm{k})\sigma_{y}-\delta_{t_{z}} \varepsilon_{z}(\bm{k})\sigma_{z},
\end{align}
where
\begin{align}
\label{eq:deltatx}\delta_{t_{x}}=&\frac{g^{2}}{2}\int_{\bm{q},\bm{p}}\frac{\varepsilon_{x}^{2}t_{y}^{2}(q_{x}^{4}+6q_{x}^{2}q_{y}^{2}+q_{y}^{4})-2\varepsilon_{y}^{2}t_{y}^{2}(q_{x}^{4}+q_{y}^{4})-(2-t_{y}^{2}/t_{x}^{2})\varepsilon_{x}^{2}\varepsilon_{z}^{2}+\varepsilon_{z}^{4}}{(\varepsilon_{x}^{2}+\varepsilon_{y}^{2}+\varepsilon_{z}^{2})^{5/2}(a(q_{x}^{2}+q_{y}^{2})+(q_{z}^{2}+p^{2})/a)},\\
\label{eq:deltaty}\delta_{t_{y}}=&\frac{g^{2}}{2}\int_{\bm{q},\bm{p}}\frac{-\varepsilon_{x}^{2}t_{x}^{2}(q_{x}^{4}+6q_{x}^{2}q_{y}^{2}+q_{y}^{4})+2\varepsilon_{y}^{2}t_{x}^{2}(q_{x}^{4}+q_{y}^{4})-(2-t_{x}^{2}/t_{y}^{2})\varepsilon_{y}^{2}\varepsilon_{z}^{2}+\varepsilon_{z}^{4}}{(\varepsilon_{x}^{2}+\varepsilon_{y}^{2}+\varepsilon_{z}^{2})^{5/2}(a(q_{x}^{2}+q_{y}^{2})+(q_{z}^{2}+p^{2})/a)},\\
\label{eq:deltatz}\delta_{t_{z}}=&\frac{g^{2}}{2}\int_{\bm{q},\bm{p}}\frac{(\varepsilon_{x}^{2}+\varepsilon_{y}^{2})(\varepsilon_{x}^{2}+\varepsilon_{y}^{2}-\varepsilon_{z}t_{z}(5q_{z}^{2}-p^{2}))}{(\varepsilon_{x}^{2}+\varepsilon_{y}^{2}+\varepsilon_{z}^{2})^{5/2}}\frac{1}{a(q_{x}^{2}+q_{y}^{2})+(q_{z}^{2}+p^{2})/a}.
\end{align}
To prove $t_x=t_y$ at low energies, let us define $T=t_x/t_y$. Then, the beta function of $T$ is given by
\begin{align}
\frac{1}{T}\frac{dT}{d\ell}=&\frac{\delta_{t_{x}}-\delta_{t_{y}}}{\ell}.
\end{align}
From Eqs.~(\ref{eq:deltatx}) and (\ref{eq:deltaty}), $\delta_{t_{x}}-\delta_{t_{y}}$ is given by
\begin{align}
\delta_{t_{x}}-\delta_{t_{y}}=&\frac{g^{2}}{2}\int_{\bm{q},\bm{p}}\frac{(t_{x}^{2}+t_{y}^{2})\varepsilon_{x}^{2}(q_{x}^{4}+6q_{x}^{2}q_{y}^{2}+q_{y}^{4})-2(t_{x}^{2}+t_{y}^{2})\varepsilon_{y}^{2}(q_{x}^{4}+q_{y}^{4})-((2-t_{y}^{2}/t_{x}^{2})\varepsilon_{x}^{2}-(2-t_{x}^{2}/t_{y}^{2})\varepsilon_{y}^{2})\varepsilon_{z}^{2}}{(\varepsilon_{x}^{2}+\varepsilon_{y}^{2}+\varepsilon_{z}^{2})^{5/2}(a(q_{x}^{2}+q_{y}^{2})+(q_{z}^{2}+p^{2})/a)}\nonumber\\
=&\frac{g^{2}}{2t_{y}}\int_{\bm{q},\bm{p}}\frac{(1+T^{2})((q_{x}^{2}-q_{y}^{2})^{2}(q_{x}^{4}+6q_{x}^{2}q_{y}^{2}+q_{y}^{4})T^{2}-8q_{x}^{2}q_{y}^{2})}{(T^{2}(q_{x}^{2}-q_{y}^{2})^{2}+4q_{x}^{2}q_{y}^{2}+\beta^{2}(q_{z}^{2}+p^{2})^{2})^{5/2}(a(q_{x}^{2}+q_{y}^{2})+(q_{z}^{2}+p^{2})/a)}\nonumber\\
&+\frac{g^{2}}{2t_{y}}\int_{\bm{q},\bm{p}}\frac{\beta^{2}(q_{z}^{2}+p^{2})^{2}(q_{x}^{4}+6q_{x}^{2}q_{y}^{2}+q_{y}^{4}-2(q_{x}^{4}+q_{y}^{4})T^{2})}{(T^{2}(q_{x}^{2}-q_{y}^{2})^{2}+4q_{x}^{2}q_{y}^{2}+\beta^{2}(q_{z}^{2}+p^{2})^{2})^{5/2}(a(q_{x}^{2}+q_{y}^{2})+(q_{z}^{2}+p^{2})/a)},
\end{align}
where $\beta\equiv t_{z}/t_{y}$.\\
Expanding $\delta_{t_{x}}-\delta_{t_{y}}$ in terms of $\delta T=T-1$, then we have
\begin{align}
\delta_{t_{x}}-\delta_{t_{y}}\approx&\frac{g^{2}}{2t_{y}}\int_{\bm{q},\bm{p}}\frac{(q_{x}^{4}-6q_{x}^{2}q_{y}^{2}+q_{y}^{4})(2(q_{x}^{2}+q_{y}^{2})^{2}-(q_{z}^{2}+p^{2})^{2}\beta^{2})}{((q_{x}^{2}+q_{y}^{2})^{2}+\beta^{2}(q_{z}^{2}+p^{2})^{2})^{5/2}(a(q_{x}^{2}+q_{y}^{2})+(q_{z}^{2}+p^{2})/a)}\nonumber\\
&-\frac{g^{2}}{2t_{y}}\delta T\int_{\bm{q},\bm{p}}\frac{4(q_{x}^{2}+q_{y}^{2})^{2}(q_{x}^{8}-22q_{x}^{6}q_{y}^{2}+50q_{x}^{4}q_{y}^{4}-22q_{x}^{2}q_{y}^{6}+q_{y}^{8})}{((q_{x}^{2}+q_{y}^{2})^{2}+\beta^{2}(q_{z}^{2}+p^{2})^{2})^{7/2}(a(q_{x}^{2}+q_{y}^{2})+(q_{z}^{2}+p^{2})/a)} \nonumber \\
&+\frac{g^{2}}{2t_{y}}\delta T\int_{\bm{q},\bm{p}}\frac{\beta^{2}(q_{z}^{2}+p^{2})^{2}(7q_{x}^{8}-40q_{x}^{6}q_{y}^{2}+2q_{x}^{4}q_{y}^{4}-40q_{x}^{2}q_{y}^{6}+7q_{y}^{8})}{((q_{x}^{2}+q_{y}^{2})^{2}+\beta^{2}(q_{z}^{2}+p^{2})^{2})^{7/2}(a(q_{x}^{2}+q_{y}^{2})+(q_{z}^{2}+p^{2})/a)}\nonumber\\
&-\frac{g^{2}}{2t_{y}}\delta T\int_{\bm{q},\bm{p}}\frac{4\beta^{4}(q_{x}^{4}+q_{y}^{4})(q_{z}^{2}+p^{2})^{4}}{((q_{x}^{2}+q_{y}^{2})^{2}+\beta^{2}(q_{z}^{2}+p^{2})^{2})^{7/2}(a(q_{x}^{2}+q_{y}^{2})+(q_{z}^{2}+p^{2})/a)}\nonumber\\
=&-\frac{A_{d-2}g^{2}}{12\pi \sqrt{t_{y}t_{z}}\Lambda^{4-d}}\left(\frac{9a^{5}\beta^{5/2}}{4}\int_{0}^{\infty} {dr}\frac{r^{5}(4a^{4}\beta^{2}-r^{4})}{(r^{4}+a^{4}\beta^{2})^{7/2}(r^{2}+1)}\right)\delta T\ell\nonumber\\
=&-\alpha G_{T}(\gamma)\delta T\ell,
\end{align}
where $\alpha=\frac{A_{d-2}g^{2}}{\sqrt{t_{\perp}t_{z}}\Lambda^{4-d}}$, 
$\gamma=\frac{a\sqrt{\beta}}{2}$, and $A_{d}=\frac{1}{6\pi(4\pi)^{d/2}\Gamma(d/2)}$. Here, we introduce the function $G_T(x)$ defined by
\begin{align}
G_{T}(x)=&72x^{5}\int_{0}^{\infty} {dr}\frac{r^{5}(64x^{4}-r^{4})}{(r^{4}+16x^{4})^{7/2}(r^{2}+1)}\nonumber\\
=&\frac{3x}{4(1+16x^{4})^{7/2}}\left[	\sqrt{1+16x^{4}}(1+160x^{4}-832x^{6}-1536x^{8}+2048x^{10})\right. \nonumber \\
&\quad\quad\quad\quad\quad\quad\quad\quad\left.+48x^{4}(-1+64x^{4})\ln\left(\frac{4x^{2}(4x^{2}+\sqrt{1+16x^{4}})}{-1+\sqrt{1+16x^{4}}}\right)	\right].
\end{align}
Note that $G_{T}(\gamma)$ is positive for all $\gamma$. Then, we see 
\begin{align}
\frac{1}{\delta T}\frac{d\delta T}{d\ell}\approx&-\alpha G_{T}(\gamma)\delta T
\end{align}
is negative (positive) for positive (negative) $\delta T$.
Therefore, $\delta T$ flows to 0, which means $T=1$ is a stable fixed point and we arrive at the conclusion that $t_{x}=t_{y}\equiv t_{\perp}$ at the low energies. Combining the results of Secs.~\ref{app:aa} and \ref{app:txtytr}, we can use the following form of action at low energies,
\begin{align}
\mathcal{S}=\int d\tau d^{d}x\;\left[ \psi^{\dagger}(\partial_{\tau}-ig\phi+\hat{\mathcal{H}}_{0}(-i\nabla))\psi+\frac{1}{2}\left(a\left\{(\partial_{x}\phi)^{2}+(\partial_{y}\phi)^{2}\right\}+\frac{1}{a}(\partial_{z}\phi)^{2}\right)\right],
\end{align}
where
\begin{align}
\mathcal{H}_{0}(\bm{k})=&t_{\perp}(k_{x}^{2}-k_{y}^{2})\sigma_{x}+2t_{\perp}k_{x}k_{y}\sigma_{y}+t_{z}k_{z}^{2}\sigma_{z}.\label{eq:action}
\end{align}

\subsection{Renormalization group equations in the $\epsilon=4-d$ expansion}
\begin{figure}[t]
\centering
\includegraphics[width=0.45\linewidth]{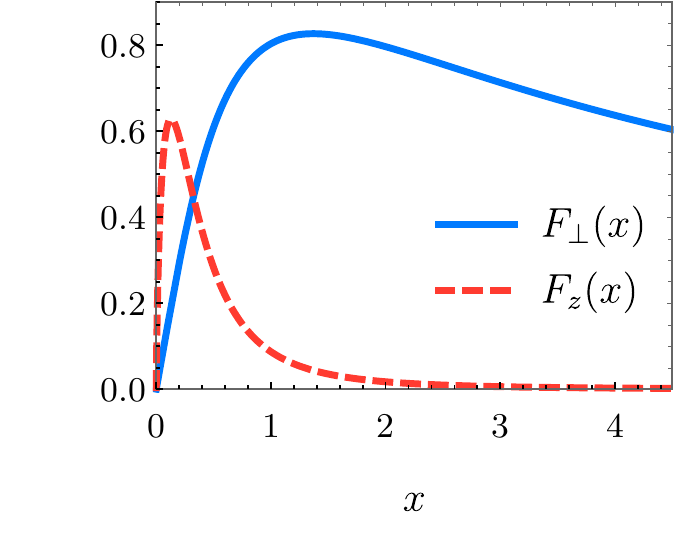}
\caption{Plots of $F_{\perp}(x)$ and $F_{z}(x)$. The blue solid line and red dashed line represent $F_{\perp}(x)$ and $F_{z}(x)$, respectively.}\label{fig:fpfz}
\end{figure}

In this section, we will show the details of the RG analysis using the $\epsilon=4-d$ expansion. From  Eqs.~(\ref{eq:Pi}) and (\ref{eq:deltatx})$-$(\ref{eq:deltatz}) 
with $t_{x}=t_{y}=t_{\perp}$ and $a_{x}=a_{y}=a_{z}^{-1}=a$, we obtain the fermion and boson self-energies, respectively, given by
\begin{align}
\label{eq:fermion self-energy}\Sigma(i\Omega,\bm{q})=&(-ig)^{2}\int_{\omega,\bm{k},\bm{p}}\!\! G_{0}(i\omega+i\Omega,\bm{k}+\bm{q})D_{0}(i\omega,\bm{k}) \nonumber\\
\approx&-\alpha F_{\perp}(\gamma)\ell\left[t_{\perp}(q_{x}^{2}-q_{y}^{2})\sigma_{x}+2t_{\perp}q_{x}q_{y}\sigma_{y} \right]-\alpha F_{z}(\gamma)\ell\left(t_{z}q_{z}^{2}\right)\sigma_{z},\\
\label{eq:boson self-energy}\Pi(\bm{q})=&g^{2}\int_{\omega,\bm{k},\bm{p}}\!\! \text{Tr}\left[G_{0}(i\omega,\bm{k}+\bm{q}/2)G_{0}(i\omega,\bm{k}-\bm{q}/2)\right] \nonumber \\
\approx&-N_{f}\alpha\left[	\frac{a}{\gamma} q_{\perp}^2+\frac{\gamma}{a}q_{z}^{2}	\right]\ell,
\end{align}
where $F_{\perp}(\gamma)$ and $F_{z}(\gamma)$ are given by
\begin{align}
F_{\perp}(x)\equiv&\left.\frac{\delta_{x}}{\alpha \ell}\right|_{t_{x}=t_{y}=t_{\perp},\;\; a_{x}=a_{y}=a_{z}^{-1}=a} \nonumber
\\=&48x^{5}\int_{0}^{\infty} dr\;\frac{ r(32x^{4}-r^{4})}{(r^{4}+16x^{4})^{5/2}(r^{2}+1)}\nonumber
\\=&\frac{3x}{2(1+16x^{4})^{5/2}}\left[	\sqrt{1+16x^{4}}(1+64x^{4}-192x^{6})-16x^{4}(1-32x^{4})\ln\left(	\frac{4x^{2}\left(4x^{2}+\sqrt{1+16x^{4}}\right)}{-1+\sqrt{1+16x^{4}}}	\right)	\right],
\\
F_{z}(x)\equiv&\left.\frac{\delta_{z}}{\alpha \ell}\right|_{t_{x}=t_{y}=t_{\perp},\;\;a_{x}=a_{y}=a_{z}^{-1}=a}\nonumber
\\=&6x\int_{0}^{\infty} dr\frac{r^{5}(r^{4}-32x^{4})}{(r^{4}+16x^{4})^{5/2}(r^{2}+1)}\nonumber
\\=&\frac{3x}{(1+16x^{4})^{5/2}}\left[	\sqrt{1+16x^{4}}(-2+12x^{2}+16x^{4})+(1-32x^{4})\ln\left(	\frac{4x^{2} \left(4x^{2}+\sqrt{1+16x^{4}}\right)}{-1+\sqrt{1+16x^{4}}}	\right)	\right].
\end{align}
Figure \ref{fig:fpfz} shows the plots of $F_{\perp}(x)$ and $F_{z}(x)$. Then, after rescaling $z\rightarrow ze^{\ell}$, $(x,y)\rightarrow (x,y)e^{z_{\perp}\ell}$, and $\tau\rightarrow e^{z\ell}\tau$, and introducing the renormalization constant, $\psi\rightarrow \psi/Z_{\psi}^{1/2}$, $\phi\rightarrow \phi/Z_{\phi}^{1/2}$, $t_{\perp}\rightarrow t_{\perp}/Z_{t_{\perp}}$, $t_{z}\rightarrow t_{z}/Z_{t_{z}}$, $a\rightarrow a/Z_{a}$, and $g\rightarrow g/Z_{g}$, we arrive at the following renormalized action,
\begin{align}
\mathcal{S}_{\text{renorm}}=
&\int d\tau d^{d}x \left[ \psi^{\dagger}\bigg(\partial_{\tau}-ig\phi+\mathcal{H}_{0}(-i\nabla)-\Sigma(-i\nabla)\bigg)\psi+\frac{1}{2}\left(a\left\{(\partial_{x}\phi)^{2}+(\partial_{y}\phi)^{2}\right\}+\frac{1}{a}(\partial_{z}\phi)^{2}\right)-\frac{1}{2}\phi\Pi(-i\nabla)\phi\right]\nonumber\\
=&\int d\tau d^{d}x\;\frac{e^{(z+2z_{\perp}+d-2)\ell}}{Z_{\psi}}\psi^{\dagger}\left[e^{-z\ell}\partial_{\tau}-\frac{1}{Z_{g}Z_{\phi}^{1/2}}ig\phi\right.\\
&\left.+\frac{e^{-2z_{\perp}\ell}}{Z_{t_{\perp}}}\left(	1+\alpha F_{\perp}(\gamma)\ell\right)t_{\perp}\left((\partial_{y}^{2}-\partial_{x}^{2})\sigma_{x}-2\partial_{x}\partial_{y}\sigma_{y}\right)-\frac{e^{-2\ell}}{Z_{t_{z}}}(1+\alpha F_{z}(\gamma))t_{z}\partial_{z}^{2}\sigma_{z}\right]\psi\nonumber\\
&+\int d\tau d^{3}x\frac{e^{(z+2z_{\perp}+d-2)\ell}}{2Z_{\phi}}\left[\frac{e^{-2z_{\perp}\ell}}{Z_{a}}2\left(	1+N_{f}\frac{\alpha}{\gamma}\ell	\right)a((\partial_{x}\phi)^{2}+(\partial_{y}\phi))+e^{-2\ell}Z_{a}\left(	1+N_{f}\alpha\gamma\ell	\right)\frac{1}{a}(\partial_{z}\phi)^{2}\right].\nonumber
\end{align}

Requiring the scaling invariance of the action, we obtain the renormalization constants as follows:
\begin{align}
Z_{\psi}=&1+\left[2z_{\perp}+(d-2)\right]\ell,\\
Z_{t_{\perp}}=&1+\left[z-2z_{\perp}+\alpha F_{\perp}(\gamma)\right]\ell,\\
Z_{t_{z}}=&1+\left[z-2+\alpha F_{z}(\gamma)\right]\ell,\\
Z_{\phi}=&1+\left[	z+z_{\perp}+(d-3)+\frac{N_{f}\alpha}{2}\left(\frac{1}{\gamma}+\gamma\right)	\right]\ell,\\
Z_{a}=&1+\left[1-z_{\perp}+\frac{N_{f}\alpha}{2}\left(\frac{1}{\gamma}-\gamma\right)\right]	\ell,\\
Z_{g}=&1+\left[\frac{z-z_{\perp}-(d-3)}{2}-\frac{N_{f}\alpha}{4}\left(\frac{1}{\gamma}+\gamma\right)		\right]\ell.
\end{align}
From these renormalization constants, we can obtain the following RG equations for $d=4-\epsilon$,
\begin{align}
\frac{1}{t_{\perp}}\frac{dt_{\perp}}{d\ell}=&z-2z_{\perp}+\alpha F_{\perp}(\gamma),\\
\frac{1}{t_{z}}\frac{dt_{z}}{d\ell}=&z-2+\alpha F_{z}(\gamma),\\
\frac{1}{a}\frac{da}{d\ell}=&1-z_{\perp}+\frac{N_{f}\alpha}{2}\left(\frac{1}{\gamma}-\gamma\right),\\
\frac{1}{g^{2}}\frac{dg^{2}}{d\ell}=&z-z_{\perp}-1+\epsilon-\frac{N_{f}\alpha}{2}\left(\frac{1}{\gamma}+\gamma	\right).
\end{align}
Thus, we find the RG equations for the dimensionless parameters $\alpha$ and $\gamma$ as follows:
\begin{align}
\frac{1}{\alpha}\frac{d\alpha}{d\ell}=&\epsilon-\frac{N_{f}\alpha}{2}\left(\frac{1}{\gamma}+\gamma	\right)
-\frac{\alpha}{2}\Big(	F_{z}(\gamma)+F_{\perp}(\gamma)	\Big),\\
\frac{1}{\gamma}\frac{d\gamma}{d\ell}=&\frac{N_{f}\alpha}{2}\left(\frac{1}{\gamma}-\gamma\right)
+\frac{\alpha}{2}\Big(F_{z}(\gamma)-F_{\perp}(\gamma)\Big).
\end{align}

\subsection{Effects of the symmetry-allowed parabolic term}
If we include the symmetry-allowed parabolic term, $s_{\perp} (k_{x}^{2}+k_{y}^{2})\sigma_z$, the non-interacting Hamiltonian $\mathcal{H}_0$ is modified as
\begin{align}
\mathcal{H}_{0}=&t_{\perp}(k_{x}^{2}-k_{y}^{2})\sigma_{x}+2t_{\perp}k_{x}k_{y}\sigma_{y}+\left[Bt_{z}k_{z}^{2}+s_{\perp}(k_{x}^{2}+k_{y}^{2})\right]\sigma_{z},
\end{align}
where $B=\pm1$ for the topologically trivial and nontrivial insulator phases, respectively.\\

\subsubsection{Boson self-energy\label{sec:app:bosonselfenergy}}
Similarly as in Eq.~(\ref{eq:boson self-energy}), we can obtain the boson self-energy in the presence of the symmetry-allowed parabolic term as
\begin{align}
\Pi(i\Omega,\bm{q})=&-N_{f}(ig)^{2}\int_{\omega,\bm{k},\bm{p}} \text{Tr}[G_{0}(i\Omega+i\omega,\bm{k}+\bm{q})G_{0}(i\omega,\bm{k})]\nonumber\\
\approx&-N_{f}\alpha\left[\frac{1}{\gamma}\left(\frac{2+\lambda^{2}}{2}-B\frac{\lambda(5+2\lambda^{2})}{4\sqrt{1+\lambda^{2}}}\right)a q_{\perp}^{2}+\gamma\left(\frac{1+2\lambda^{2}}{\sqrt{1+\lambda^{2}}}- 2B\lambda\right)\frac{1}{a}q_{z}^{2}	\right],
\end{align}
where $\lambda=\frac{s_{\perp}}{t_{\perp}}$. 

\subsubsection{Fermion self-energy\label{sec:app:fermionselfenergy}}
Similarly as in Eq.~(\ref{eq:fermion self-energy}), we can obtain the fermion self-energy as
\begin{align}
\Sigma(i\omega,\bm{k})=&(ig)^{2}\int_{\Omega,\bm{q},\bm{p}} G_{0}(i\omega+i\Omega,\bm{k}+\bm{q})D_{0}(i\Omega,\bm{q})\nonumber\\
\approx&-\delta_{t_{\perp}} \left[ t_{\perp}(k_{x}^{2}-k_{y}^{2})\sigma_{x}+2t_{\perp}k_{x}k_{y}\sigma_{y}\right]-\left[\delta_{t_{z}}Bt_{z}k_{z}^{2}+\delta_{s_{\perp}}s_{\perp}(k_{x}^{2}+k_{y}^{2})\right]\sigma_{z},
\end{align}
where $\delta_{t_{\perp}}$, $\delta_{t_z}$ and $\delta_{s_{\perp}}$ are, respectively, given by
\begin{align}
\delta_{t_{\perp}}=&\frac{g^{2}}{2}\int_{\bm{k},\bm{p}}\frac{t_{z}^{2}k_{\perp}k_{z}^{d+1}( 2(Bt_{z}k_{z}^{2}+s_{\perp}k_{\perp}^{2})^{2}-t_{\perp}^{2}k_{\perp}^{4})}{2\big(t_{\perp}^{2}k_{\perp}^{4}+(Bt_{z}k_{z}^{2}+s_{\perp} k_{\perp}^{2})^{2}\big)^{5/2}(a k_{\perp}^{2}+\frac{1}{a}k_{z}^{2})}\nonumber\\
=&\frac{A_{d-2}g^{2}\;\ell}{\sqrt{t_{\perp}t_{z}}\Lambda^{4-d}}\int dr\;\frac{3(2\gamma)^{5}r(-r^{4}+2(4B\gamma^{2}+\lambda r^{2})^{2})}{2(1+r^{2})(r^{4}+\big(4B\gamma^{2}+\lambda r^{2})^{2}\big)^{5/2}}\nonumber\\
=&\alpha F_{\perp}(\gamma,\lambda)\ell,\\
\delta_{t_{z}}=&\frac{g^{2}}{2}\int_{\bm{k},\bm{p}}\frac{t_{\perp}^{3}k_{\perp}^{5}k_{z}^{d-3}(t_{\perp}^{2}k_{\perp}^{4}-(2Bt_{z}k_{z}^{2}-s_{\perp} k_{\perp}^{2})(Bt_{z}k_{z}^{2}+s_{\perp} k_{\perp}^{2}))}{\big(t_{\perp}^{2}k_{\perp}^{4}+(Bt_{z}k_{z}^{2}+s_{\perp} k_{\perp}^{2})^{2}\big)^{5/2}(a k_{\perp}^{2}+\frac{1}{a}k_{z}^{2})}\nonumber\\
=&\frac{A_{d-2}g^{2}\;\ell}{\sqrt{t_{\perp}t_{z}}\Lambda^{4-d}}\int dr\;\frac{6\gamma r^{5}(r^{4}-(8B\gamma^{2}-\lambda r^{2})(2B\gamma^{2}+\lambda r^{2}))}{(1+r^{2})\big(r^{4}+(4B\gamma^{2}+\lambda r^{2})^{2}\big)^{5/2}}\nonumber\\
=&\alpha F_{z}(\gamma,\lambda)\ell,\\
\delta_{s_{\perp}}=&\frac{g^{2}}{2}\int_{\bm{k},\bm{p}}\frac{B}{s_{\perp}}\frac{t_{\perp}^{2}t_{z}k_{\perp}^{3}k_{z}^{d-1}(t_{\perp}^{2}k_{\perp}^{4}-(2Bt_{z}k_{z}^{2}-s_{\perp} k_{\perp}^{2})(Bt_{z}k_{z}^{2}+s_{\perp} k_{\perp}^{2}))}{2\big(t_{\perp}^{2}k_{\perp}^{4}+(Bt_{z}k_{z}^{2}+s_{\perp} k_{\perp}^{2})^{2}\big)^{5/2}(a k_{\perp}^{2}+\frac{1}{a} k_{z}^{2})}\nonumber\\
=&\frac{B}{\lambda}\frac{A_{d-2}g^{2}\;\ell}{\sqrt{t_{\perp}t_{z}}\Lambda^{4-d}}\int dr\;\frac{3(2\gamma)^{3} r^{3}(r^{4}-(8B\gamma^{2}-\lambda r^{2})(4B\gamma^{2}+\lambda r^{2}))}{(1+r^{2})\big(r^{4}+(4B\gamma^{2}+\lambda r^{2})^{2}\big)^{5/2}}\nonumber\\
=&\frac{B}{\lambda}\alpha F_{s}(\gamma,\lambda)\ell.
\end{align}
Here, we introduce the following dimensionless functions,
\begin{align}
F_{\perp}(\gamma,\lambda)=&\int_{0}^{\infty} dr\;\frac{48\gamma^{5} r(-r^{4}+2(4B\gamma^{2}+\lambda r^{2})^{2})}{(1+r^{2})(r^{4}+(4B\gamma^{2}+\lambda r^{2})^{2})^{5/2}}\nonumber\\
=&\frac{3\gamma}{2(1+(4B\gamma^{2}-\lambda)^{2})^{2}}\left[	(1+\lambda^{2})^{3/2}+64\gamma^{4}(-3\gamma^{2}+\sqrt{1+\lambda^{2}})-4B\gamma^{2}\lambda(-12\gamma^{2}+5\sqrt{1+\lambda^{2}})	\right.\nonumber\\
\phantom{=}&\left.-\frac{16\gamma^{4}(1-2(4B\gamma^{2}-\lambda)^{2})}{\sqrt{1+(4B\gamma^{2}-\lambda)^{2}}}\ln\left(\frac{4\gamma^{2}\Big(4\gamma^{2}-B\lambda+\sqrt{1+(4B\gamma^{2}-\lambda)^{2}}\Big)}{-1+4B\gamma^{2}\lambda-\lambda^{2}+\sqrt{1+\lambda^{2}}\sqrt{1+(4B\gamma^{2}-\lambda)^{2}}}\right)	\right],
\\ 
F_{z}(\gamma,\lambda)=&\int_{0}^{\infty} dr\;\frac{6\gamma r^{5}(r^{4}-(8B\gamma^{2}-\lambda r^{2})(4B\gamma^{2}+\lambda r^{2}))}{(1+r^{2})(r^{4}+(4B\gamma^{2}+\lambda r^{2})^{2})^{5/2}}\nonumber\\
=&\frac{3\gamma}{(1+(4B\gamma^{2}-\lambda))^{2}}\left[-\Big(16B\gamma^{4}\lambda-4\gamma^{2}(3+2\lambda^{2})+B\lambda(1+\lambda^{2})\Big)+\sqrt{1+\lambda^{2}}(-2+(4B\gamma^{2}-\lambda)^{2})\right.\nonumber\\
\phantom{=}&\left.+\frac{1-32\gamma^{4}+4B\gamma^{2}\lambda+\lambda^{2}}{\sqrt{1+(4B\gamma^{2}-\lambda)^{2}}}\ln\left(\frac{4\gamma^{2}\Big(4\gamma^{2}-B\lambda+\sqrt{1+(4B\gamma^{2}-\lambda)^{2}}\Big)}{-1+4B\gamma^{2}\lambda-\lambda^{2}+\sqrt{1+\lambda^{2}}\sqrt{1+(4B\gamma^{2}-\lambda)^{2}}}\right)\right],
\\
F_{s}(\gamma,\lambda)=&\int_{0}^{\infty} dr\;\frac{24\gamma^{3} r^{3}(r^{4}-(8B\gamma^{2}-\lambda r^{2})(4B\gamma^{2}+\lambda r^{2}))}{(1+r^{2})(r^{4}+(4B\gamma^{2}+\lambda r^{2})^{2})^{5/2}}\nonumber\\
=&\frac{-12\gamma}{(1+(4B\gamma^{2}-\lambda))^{2}}\left[-\Big(16B\gamma^{4}\lambda-4\gamma^{2}(3+2\lambda^{2})+B\lambda(1+\lambda^{2})\Big)+\sqrt{1+\lambda^{2}}(-2+(4B\gamma^{2}-\lambda)^{2})\right.\nonumber\\
\phantom{=}&\left.-\frac{1-32\gamma^{4}+4B\gamma^{2}\lambda+\lambda^{2}}{\sqrt{1+(4B\gamma^{2}-\lambda)^{2}}}\ln\left(\frac{4\gamma^{2}\Big(4\gamma^{2}-B\lambda+\sqrt{1+(4B\gamma^{2}-\lambda)^{2}}\Big)}{-1+4B\gamma^{2}\lambda-\lambda^{2}+\sqrt{1+\lambda^{2}}\sqrt{1+(4B\gamma^{2}-\lambda)^{2}}}\right)\right]\nonumber\\
=&-4\gamma^{2}F_{z}(\gamma,\lambda).
\end{align}
Note that in the limit $\lambda=0$, $F_{\perp}(\gamma,\lambda)=F_{\perp}(\gamma)$, and $F_{z}(\gamma,\lambda)=F_{z}(\gamma)$.

\subsubsection{RG flow equation}
From Sec.~\ref{sec:app:bosonselfenergy} and Sec.~\ref{sec:app:fermionselfenergy}, we can obtain the following RG flow equations,
\begin{align}
\frac{1}{t_{\perp}}\frac{dt_{\perp}}{d\ell}=&z-2z_{\perp}+\alpha F_{\perp}(\gamma,\lambda),\\
\frac{1}{t_{z}}\frac{dt_{z}}{d\ell}=&z-2+\alpha F_{z}(\gamma,\lambda),\\
\frac{1}{s_{\perp}}\frac{ds_{\perp}}{d\ell}=&z-2z_{\perp}-4b\frac{\alpha}{\lambda} \gamma^{2}F_{z}(\gamma,\lambda),\\
\frac{1}{a}\frac{d a}{d\ell}=&1-z_{\perp}+\frac{N_{f}\alpha}{2}\left[\frac{1}{\gamma}\left(\frac{2+\lambda^{2}}{2}-B\frac{\lambda(5+2\lambda^{2})}{4\sqrt{1+\lambda^{2}}}\right)-\gamma\left(\frac{1+2\lambda^{2}}{\sqrt{1+\lambda^{2}}}- 2B\lambda\right)\right],\\
\frac{1}{g^2}\frac{d g^2}{d\ell}=&z-z_{\perp}-1+\epsilon-\frac{N_{f} \alpha}{2}\left[\frac{1}{\gamma}\left(\frac{2+\lambda^{2}}{2}-B\frac{\lambda(5+2\lambda^{2})}{4\sqrt{1+\lambda^{2}}}\right)+\gamma\left(\frac{1+2\lambda^{2}}{\sqrt{1+\lambda^{2}}}- 2B\lambda\right)\right].
\end{align}

Then, the RG equations for the dimensionless parameters, $\alpha$, $\gamma$ and $\lambda$ are given by
\begin{align}
\frac{1}{\alpha}\frac{d\alpha}{d\ell}=&\epsilon-\frac{\alpha}{2}\left[ N_{f}\left\{\frac{1}{\gamma}\left(\frac{2+\lambda^{2}}{2}-B\frac{\lambda(5+2\lambda^{2})}{4\sqrt{1+\lambda^{2}}}\right)+\gamma\left(\frac{1+2\lambda^{2}}{\sqrt{1+\lambda^{2}}}- 2B\lambda\right)\right\}+F_{z}(\gamma,\lambda)+F_{\perp}(\gamma,\lambda)\right],\\
\frac{1}{\gamma}\frac{d\gamma}{d\ell}=&\frac{\alpha}{2}\left[N_{f}\left\{\frac{1}{\gamma}\left(\frac{2+\lambda^{2}}{2}-B\frac{\lambda(5+2\lambda^{2})}{4\sqrt{1+\lambda^{2}}}\right)-\gamma\left(\frac{1+2\lambda^{2}}{\sqrt{1+\lambda^{2}}}- 2B\lambda\right)\right\}+F_{z}(\gamma,\lambda)-F_{\perp}(\gamma,\lambda)\right],\\
\frac{1}{\lambda}\frac{d\lambda}{d\ell}=&-\frac{\alpha}{\lambda}\left[4B\gamma^{2}F_{z}(\gamma,\lambda)+\lambda F_{\perp}(\gamma,\lambda)\right].
\end{align}

For given $N_{f}$, the RG equations have unstable fixed point, $\alpha^{*}=0$ with arbitrary $\gamma^{*}$ and $\lambda^{*}$, and stable interacting fixed point, $(\alpha^{*},\gamma^{*},\lambda^{*})=(0.342\epsilon/N_{f}, 0.799-0.079/N_{f}, -B(0.875+0.032/N_{f}))$ for large $N_{f}$. Then, near the interacting fixed point,
\begin{align}
\left.\frac{1}{a}\frac{d a}{d\ell}\right|_{\text{f.p.}}&=\frac{N_{f}\alpha^{*}}{2}\left[\frac{1}{\gamma^{*}}\left(\frac{2+{\lambda^{*}}^{2}}{2}-B\frac{{\lambda^{*}}(5+2{\lambda^{*}}^{2})}{4\sqrt{1+{\lambda^{*}}^{2}}}\right)-\gamma^{*}\left(\frac{1+2{\lambda^{*}}^{2}}{\sqrt{1+{\lambda^{*}}^{2}}}- 2B{\lambda^{*}}\right)\right]>0,\\
\left.\frac{1}{\beta^{-1}}\frac{d \beta^{-1}}{d\ell}\right|_{\text{f.p.}}&=-\alpha^{*}\Big(F_{z}(\gamma^{*},\lambda^{*})-F_{\perp}(\gamma^{*},\lambda^{*})\Big)>0.
\end{align}
Thus, the bosonic and fermionic anisotropy parameters $a$ and $\beta^{-1}$ diverge at the stable interacting fixed point. Therefore, even if we keep $s_{\perp}(k_{x}^{2}+k_{y}^{2})\sigma_{z}$, the interacting fixed point still exhibits anisotropic non-Fermi liquid behaviors.

\section{Details of the large $N_{f}$ calculation}
In this section, we will show the detailed calculations of the large $N_{f}$ method.

\subsection{Boson self-energy\label{sec:LargeN-Coulomb}}
Consider the self-energy of the Coulomb interaction given by
\begin{align}
\Pi(i\Omega,\bm{q})=&-N_{f}(-ig)^{2}\int_{\omega,\bm{k}}\text{Tr}[G_{0}(i\Omega+i\omega,\bm{k}+\bm{q})G_{0}(i\omega,\bm{k})]\nonumber\\
=&-N_{f} g^{2} \int_{\bm{k}}\frac{E_{+}+E_{-}}{(E_{+}+E_{-})^{2}+\Omega^2}\left(1-\frac{\vec{\varepsilon}_{+}\cdot\vec{\varepsilon}_{-}}{E_{+}E_{-}}\right),
\end{align}
where $\varepsilon_{i\pm}=\varepsilon_{i}(\bm{k}\pm\bm{q}/2)$ and $E_{\pm}=\sqrt{\sum_{i}{{\varepsilon}_{i\pm}^{2}}}$.
\subsubsection{$q_{\perp}$ dependence}
Let us find the $q_{\perp}$ dependence in $\Pi(i\Omega,\bm{q})$ with non-zero $i\Omega$. Because of the emergent rotational symmetry along the $k_z$-axis, we put $\bm{q}_{\perp}=q_{\perp} \hat{x}$ for simplicity. After changing the integration variables, $k_{x}\rightarrow q_{\perp} x,\, k_{y} \rightarrow q_{\perp} y, k_{z}\rightarrow(t_{\perp}/t_{z})^{1/2} q_{\perp} z$, we get
\begin{align}
\Pi(i\Omega,q_{\perp})&=-\frac{N_{f}g^{2}\left|q_{\perp}\right|}{8\pi^{3}\sqrt{t_{\perp}t_{z}}}\int\text{d}^{3}x  \frac{\sqrt{\Big((x+1)^{2}+y^{2}\Big) ^{2}+z^{4}}+\sqrt{\Big( x^{2}+y^{2}\Big)^{2}+z^{4}}}{\left[\sqrt{\Big((x+1)^{2}+y^{2}\Big)^{2}+z^{4}}+\sqrt{\Big( x^{2}+y^{2}\Big)^{2}+z^{4}}\right]^{2}+\left(\frac{\Omega}{t_{\perp}\left|q_{\perp}\right|^{2}}\right)^{2}}\nonumber \\
&\phantom{=}\times\left[ 1-\frac{\Big((x+1)^{2}-y^{2}\Big) \Big(x^{2}-y^{2}\Big) +4(x+1)xy^{2}+z^{4}}{\sqrt{\Big((x+1)^{2}+y^{2}\Big) ^{2}+z^{4}}\;\sqrt{\Big(x^{2}+y^{2}\Big)^{2}+z^{4}}}\right]\nonumber\\
&=-\frac{C_{\perp_{1}}N_{f}g^{2}}{\sqrt{t_{\perp}^{2}t_{z}}}\sqrt{t_{\perp}q_{\perp}^2}\tanh(C_{\perp_{2}}\xi_{r}),
\end{align}
where $\xi_{r}=\sqrt{\frac{t_{\perp}}{|\Omega|}}|q_{\perp}|$, $C_{\perp_{1}}=0.042$, and $C_{\perp_{2}}=1.199$. The final result is a fitting function using an ansatz obtained from $\Pi(i\Omega,q_{\perp})\propto \xi_{r}^{2}$ for $\xi_{r}\ll1$, and $\Pi(i\Omega,q_{\perp})\propto \xi_{r}$ for $\xi_{r}\gg1$.

\subsubsection{$q_{z}$ dependence}
Similarly, after changing the integration variables, $k_{\perp}\rightarrow (t_{z}/t_{\perp})^{1/2} q_{\perp} r,\, k_{z}\rightarrow q_{z} z$, we get
\begin{align}
\Pi(i\Omega,q_{z})&=-\frac{N_{f}g^{2}\left|q_{z}\right|}{4\pi^{2}t_{\perp}}\int_{0}^{\infty}\text{d}r\;r\int_{-\infty}^{\infty}\text{d}z\frac{\sqrt{r^{4}+(z+1)^{4}}+\sqrt{r^{4}+z^{4}}}{\left[\sqrt{r^{4}+(z+1)^{4}}+\sqrt{r^{4}+z^{4}}\right]^{2}+\left(\frac{\Omega}{t_{z}q_{z}^{2}}\right)^{2}} \nonumber \\
&\phantom{=}\times\left[1-\frac{r^{4}+(z+1)^{2}z^{2}}{\sqrt{r^{4}+(z+1)^{4}}\;\sqrt{r^{4}+z^{4}}}\right]\nonumber \\
&=-\frac{C_{z_{1}}N_{f}g^{2}}{\sqrt{t_{\perp}^{2}t_{z}}}\sqrt{t_z q_{z}^{2}}\tanh(C_{z_{2}}\xi_{z}),
\end{align}
where $\xi_{z}=\sqrt{\frac{t_{z}}{|\Omega|}}|q_{z}|$, $C_{z_{1}}=0.016$, and $C_{z_{2}}=1.267$. The final result is a fitting function using an ansatz obtained from $\Pi(i\Omega,q_{z})\propto \xi_{z}^{2}$ for $\xi_{z}\ll1$, and $\Pi(i\Omega,q_{z})\propto \xi_{z}$ for $\xi_{z}\gg1$.

\subsubsection{Arbitrary $q$ dependence}
\begin{figure}[t]
\centering
\subfigure{
\includegraphics[scale=1.5]{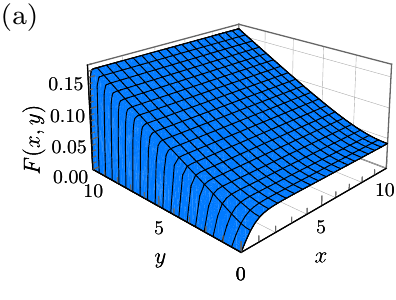}
\label{fig:Frz}}\hspace{0.95em}
\subfigure{
\includegraphics[scale=1.5]{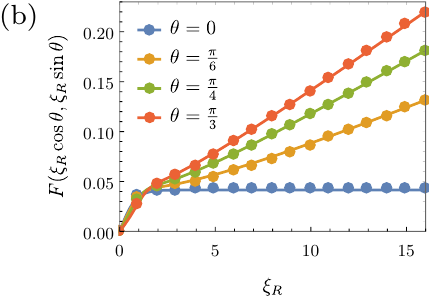}
\label{fig:Ftheta}}
\caption{(a) The dimensionless function $F(x,y)$. (b) Comparison between the exact numerical values and the ansatz $F(x,y)$. For $\sqrt{\frac{t_{\perp}}{|\Omega|}}|q_{\perp}|=\xi_{R}\cos\theta$ and $\sqrt{\frac{t_{z}}{t_{\perp}}}\left|\frac{q_{z}}{q_{\perp}}\right|=\xi_{R}\sin\theta$, we present the numerical and analytic values for $\theta=0$ (blue), $\pi/6$ (yellow), $\pi/4$ (green), and $\pi/3$ (red). The dotted points represent the exact numerical values, and the solid lines represent the values obtained by the ansatz in Eq.~(\ref{eq:ansatz}). 
The ansatz is in good agreement with the exact numerical values.
}\label{fig:Ffunctions}
\end{figure}

For arbitrary $\bm{q}$, 
\begin{align}
\Pi(i\Omega,\bm{q})&=-\frac{N_{f}g^{2}\left|q_{\perp}\right|}{8\pi^{3}\sqrt{t_{\perp}t_{Z}}}\frac{\xi_{z}}{\xi_{r}}\int\text{d}^{3}x\frac{\sqrt{\Big((x+1)^{2}+y^{2}\Big)^{2}+\frac{\xi_{z}^{4}}{\xi_{r}^{4}}(z+1)^{4}}+\sqrt{\Big(x^{2}+y^{2}\Big)^{2}+\frac{\xi_{z}^{4}}{\xi_{r}^{4}}z^{4}}}{\left[\sqrt{\Big((x+1)^{2}+y^{2}\Big)^{2}+\frac{\xi_{z}^{4}}{\xi_{r}^{4}}(z+1)^{4}}+\sqrt{\Big(x^{2}+y^{2}\Big)^{2}+\frac{\xi_{z}^{4}}{\xi_{r}^{4}}z^{4}}\right]^{2}+\left(\frac{\Omega}{t_{\perp}q_{\perp}^{2}}\right)^{2}}\nonumber \\
&\phantom{=}\times\left[1-\frac{\Big((x+1)^{2}-y^{2}\Big)\Big(x^{2}-y^{2}\Big)+4(x+1)xy^{2}-\frac{\xi_{z}^{4}}{\xi_{r}^{4}}(z+1)z}{\sqrt{\Big((x+1)^{2}+y^{2}\Big)^{2}+\frac{\xi_{z}^{4}}{\xi_{r}^{4}}(z+1)^{4}}\;\sqrt{\Big(x^{2}+y^{2}\Big)^{2}+\frac{\xi_{z}^{4}}{\xi_{r}^{4}}z^{4}}}\right] \nonumber\\
&=-\frac{N_{f}g^{2}}{\sqrt{t_{\perp}^{2}t_{z}}}\sqrt{C_{\perp_{1}}^{2}t_{\perp}q_{\perp}^{2}+C_{z_{1}}^{2}t_{z} q_{z}^{2}}\tanh\left(	\sqrt{C_{\perp_{2}}^{2}\xi_{r}^{2}+C_{z_{2}}^{2}\xi_{z}^{2}}	\right).
\end{align}
For $\Pi(i\Omega,\bm{q})\equiv-\frac{N_{f}g^{2}|q_{\perp}|}{\sqrt{t_{\perp}t_{z}}}F\left(\sqrt{\tfrac{t_{\perp}}{|\Omega|}}|q_{\perp}|,\sqrt{\tfrac{t_{z}}{t_{\perp}}}\left| \tfrac{q_{z}}{q_{\perp}}\right|\right)$, we illustrate the dimensionless function $F(x,y)$ in Fig.~\ref{fig:Ffunctions}(a) and compare it with the exact numerical values in Fig.~\ref{fig:Ffunctions}(b).

\subsection{Fermion self-energy\label{Sec:Fermion self energy in Large N}}
Using the boson self-energy obtained in Sec.~\ref{sec:LargeN-Coulomb}, we can obtain the fermion self-energy as follows:
\begin{align}
\Sigma(i\omega,\bm{k})=&(ig)^{2}\int_{\Omega,\bm{q}}G_{0}(i\Omega+i\omega,\bm{q}+\bm{k})D(i\Omega,\bm{q}) \nonumber \\
=&-g^{2}\int_{\Omega,\bm{q}}\frac{i(\Omega+\omega)+\varepsilon_{x}(\bm{k}+\bm{q})\sigma_{x}+\varepsilon_{y}(\bm{k}+\bm{q})\sigma_{y}+\varepsilon_{z}(\bm{k}+\bm{q})\sigma_{z}}{(\Omega+\omega)^{2}+\varepsilon_{x}^{2}(\bm{k}+\bm{q})+\varepsilon_{y}^{2}(\bm{k}+\bm{q})+\varepsilon_{z}^{2}(\bm{k}+\bm{q})}\frac{1}{a(q_{x}^{2}+q_{y}^{2})+q_{z}^{2}/a-\Pi(i\Omega,\bm{q})} \nonumber\\ 
\approx&-g^{2}\int_{\Omega,\bm{q}}\frac{i(\Omega+\omega)+\varepsilon_{x}(\bm{k}+\bm{q})\sigma_{x}+\varepsilon_{y}(\bm{k}+\bm{q})\sigma_{y}+\varepsilon_{z}(\bm{k}+\bm{q})\sigma_{z}}{(\Omega+\omega)^{2}+\varepsilon_{x}^{2}(\bm{k}+\bm{q})+\varepsilon_{y}^{2}(\bm{k}+\bm{q})+\varepsilon_{z}^{2}(\bm{k}+\bm{q})}\frac{1}{-\Pi(i\Omega,\bm{q})}\nonumber\\
\approx&i\omega \delta_{\omega}-\delta_{t_{\perp}}\big(\varepsilon_{x}(\bm{k})\sigma_{x}+\varepsilon_{y}(\bm{k})\sigma_{y}\big)-\delta_{t_{z}}\varepsilon_{z}(\bm{k})\sigma_{z}.
\end{align}
The corrections $\delta_{\omega}$, $\delta_{t_{\perp}}$, and $\delta_{t_{z}}$ are evaluated in the following subsections.

\subsubsection{$\omega$ correction $\delta_{\omega}$}
The correction $\delta_\omega$ is given by
\begin{align}
\delta_{\omega}=&-g^{2}\int_{\Omega,\bm{q}}\frac{t_{\perp}^{2}(q_{x}^{2}+q_{y}^{2})^{2}+t_{z}^{2}q_{z}^{4}-\Omega^{2}}{\left[t_{\perp}^{2}(q_{x}^{2}+q_{y}^{2})^{2}+t_{z}^{2}q_{z}^{4}+\Omega^{2}\right]^{2}}
\frac{\coth\left(	\frac{\sqrt{C_{\perp_{2}}^{2}(q_{x}^{2}+q_{y}^{2})+C_{z_{2}}^{2}\beta q_{z}^{2}}}{(\Omega/t_{\perp})^{1/2}}	\right)}{\frac{N_{f}e^{2}}{(t_{\perp}t_{z})^{1/2}}\sqrt{C_{\perp_{1}}^{2}(q_{x}^{2}+q_{y}^{2})+C_{z_{1}}^{2}\beta q_{z}^{2}}} \nonumber \\
=&-\frac{(t_{\perp}t_{z})^{1/2}}{8\pi^{3}N_{f}}\frac{1}{t_{\perp}^{2}}\int_{-\infty}^{\infty}d\Omega \int_{\mu<|q_{z}|<\Lambda}dq_{z}\int_{-\infty}^{\infty}dq_{\perp}\;q_{\perp}\frac{q_{\perp}^{4}+\beta^{2}q_{z}^{4}-\Omega^{2}/t_{\perp}^{2}}{\left[q_{\perp}^{4}+\beta^{2}q_{z}^{4}+\Omega^{2}/t_{\perp}^{2}\right]^{2}}
\frac{\coth\left(	\frac{\sqrt{C_{\perp_{2}}^{2}q_{\perp}^{2}+C_{z_{2}}^{2}\beta q_{z}^{2}}}{(\Omega/t_{\perp})^{1/2}}	\right)}{\sqrt{C_{\perp_{1}}^{2}q_{\perp}^{2}+C_{z_{1}}^{2}\beta q_{z}^{2}}}.
\end{align}
After changing the integration variables, $q_{\perp}\rightarrow \sqrt{\beta}q_{z}a$ and $\Omega\rightarrow \beta t_{\perp} q_{z}^{2}b$, we have
\begin{align}
\delta_{\omega}=&\frac{(t_{\perp}t_{z})^{1/2}}{N_{f}}\frac{1}{t_{\perp}^{2}}\frac{t_{\perp}\beta^{2}}{\beta^{5/2}}\ln(\Lambda/\mu)\int_{0}^{\infty}da\int_{0}^{\infty}db\;\frac{a}{2\pi^{3}}\frac{-a^{2}-1+b^{2}}{(a^{4}+1+b^{2})^{2}}\frac{\coth\left(	\sqrt{(C_{\perp_{2}}^{2}a^{2}+C_{z_{2}}^{2})/b}	\right)}{\sqrt{C_{\perp_{1}}^{2}a^{2}+C_{z_{1}}^{2}}}\nonumber \\
=&\frac{C_{\omega}}{N_{f}}\ln(\Lambda/\mu),
\end{align}
where $C_{\omega}=0.366072$. Note that $\delta_\omega$ has a logarithmic divergence both in the UV and IR cutoffs.

\subsubsection{$t_{\perp}$ correction $\delta_{t_{\perp}}$}
The correction $\delta_{t_{\perp}}$ is given by
\begin{align}
\delta_{t_{\perp}}=&g^{2}\int_{\Omega,\bm{q}}\frac{(\Omega^{2}+t_{z}^{2}q_{z}^{4})\big(\Omega^{2}-3t_{\perp}^{2}(q_{x}^{2}+q_{y}^{2})^{2}+t_{z}^{2}q_{z}^{4}\big)}{\left[\Omega^{2}+t_{\perp}^{2}(q_{x}^{2}+q_{y}^{2})^{2}+t_{z}^{2}q_{z}^{4}\right]^{3}}
\frac{\coth\left(	\frac{\sqrt{C_{\perp_{2}}^{2}(q_{x}^{2}+q_{y}^{2})+C_{z_{2}}^{2}\beta q_{z}^{2}}}{(\Omega/t_{\perp})^{1/2}}	\right)}{\frac{N_{f}e^{2}}{(t_{\perp}t_{z})^{1/2}}\sqrt{C_{\perp_{1}}^{2}(q_{x}^{2}+q_{y}^{2})+C_{z_{1}}^{2}\beta q_{z}^{2}}} \nonumber\\
=&\frac{(t_{\perp}t_{z})^{1/2}}{8\pi^{3}t_{\perp}^{2}N_{f}}\int_{-\infty}^{\infty}d\Omega \int_{\mu<|q_{z}|<\Lambda} dq_{z}\int_{-\infty}^{\infty}dq_{\perp}\;q_{\perp}\frac{(\Omega^{2}/t_{\perp}^{2}+\beta^{2}q_{z}^{4})(\Omega^{2}/t_{\perp}^{2}-3q_{\perp}^{4}+\beta^{2}q_{z}^{4})}{\left[\Omega^{2}/t_{\perp}^{2}+q_{\perp}^{4}+\beta^{2}q_{z}^{4}\right]^{3}}
\frac{\coth\left(	\frac{\sqrt{C_{\perp_{2}}^{2}q_{\perp}^{2}+C_{z_{2}}^{2}\beta q_{z}^{2}}}{(\Omega/t_{\perp})^{1/2}}	\right)}{\sqrt{C_{\perp_{1}}^{2}q_{\perp}^{2}+C_{z_{1}}^{2}\beta q_{z}^{2}}}.
\end{align}
After changing the integration variables, $q_{\perp}\rightarrow \sqrt{\beta}q_{z}a$ and $\Omega\rightarrow \beta t_{\perp} q_{z}^{2}b$, we have
\begin{align}
\delta_{t_{\perp}}=&\frac{(t_{\perp}t_{z})^{1/2}}{t_{\perp}^{2}N_{f}}\frac{t_{\perp}\beta^{2}}{\beta^{5/2}}\ln(\Lambda/\mu)\int_{0}^{\infty}da\int_{0}^{\infty}db\;\frac{a}{2\pi^{3}}\frac{(1+b^{2})(-3a^{4}+1+b^{2})}{(a^{4}+1+b^{2})^{3}}
\frac{\coth\left(	\sqrt{(C_{\perp_{2}}^{2}a^{2}+C_{z_{2}}^{2})/b}	\right)}{\sqrt{C_{\perp_{1}}^{2}a^{2}+C_{z_{1}}^{2}}}\nonumber\\
=&\frac{C_{t_{\perp}}}{N_{f}}\ln(\Lambda/\mu),
\end{align}
where $C_{t_{\perp}}=0.614362$. Note that $\delta_{t_{\perp}}$ has a logarithmic divergence both in the UV and IR cutoffs.

\subsubsection{$t_{z}$ correction $\delta_{t_{z}}$}
The correction $\delta_{t_{z}}$ is given by
\begin{align}
\delta_{t_{z}}=&g^{2}\int_{\Omega,\bm{q}}\frac{16t_{z}^{4}q_{z}^{8}+\Big(\Omega^{2}+t_{\perp}^{2}(q_{x}^{2}+q_{y}^{2})^{2}+t_{z}^{2}k_{z}^{4}\Big)\Big(\Omega^{2}+t_{\perp}^{2}(q_{x}^{2}+q_{y}^{2})^{2}-13t_{z}^{2}k_{z}^{4}\Big)}{\left[\Omega^{2}+t_{\perp}^{2}(q_{x}^{2}+q_{y}^{2})^{2}+t_{z}^{2}q_{z}^{4}\right]^{3}}\nonumber\\
&\quad\quad\quad\quad\quad\quad\times\frac{\coth\left(	\frac{\sqrt{C_{\perp_{2}}^{2}(q_{x}^{2}+q_{y}^{2})+C_{z_{2}}^{2}\beta q_{z}^{2}}}{(\Omega/t_{\perp})^{1/2}}	\right)}{\frac{N_{f}e^{2}}{(t_{\perp}t_{z})^{1/2}}\sqrt{C_{\perp_{1}}^{2}(q_{x}^{2}+q_{y}^{2})+C_{z_{1}}^{2}\beta q_{z}^{2}}}\\
=&\frac{(t_{\perp}t_{z})^{1/2}}{8\pi^{3}t_{\perp}^{2}N_{f}}\int_{-\infty}^{\infty}d\Omega \int_{\mu<|q_{z}|<\Lambda} dq_{z}\int_{-\infty}^{\infty}dq_{\perp}\;q_{\perp}\frac{16\beta^{4}q_{z}^{8}+\Big(\Omega^{2}/t_{\perp}^{2}+q_{\perp}^{4}+\beta^{2}k_{z}^{4}\Big)\Big(\Omega^{2}/t_{\perp}^{2}+q_{\perp}^{4}-13\beta^{2}k_{z}^{4}\Big)}{\left[\Omega^{2}/t_{\perp}^{2}+q_{\perp}^{4}+\beta^{2}q_{z}^{4}\right]^{3}} \nonumber \\
&\quad\quad\quad\quad\quad\quad\times\frac{\coth\left(	\frac{\sqrt{C_{\perp_{2}}^{2}q_{\perp}^{2}+C_{z_{2}}^{2}\beta q_{z}^{2}}}{(\Omega/t_{\perp})^{1/2}}	\right)}{\sqrt{C_{\perp_{1}}^{2}q_{\perp}^{2}+C_{z_{1}}^{2}\beta q_{z}^{2}}}.
\end{align}
After changing the integration variables, $q_{\perp}\rightarrow \sqrt{\beta}q_{z}a$ and $\Omega\rightarrow \beta t_{\perp} q_{z}^{2}b$, we have
\begin{align}
\delta_{t_{z}}=&\frac{(t_{\perp}t_{z})^{1/2}}{t_{\perp}^{2}N_{f}}\frac{t_{\perp}\beta^{2}}{\beta^{5/2}}\ln(\Lambda/\mu)\int_{0}^{\infty}da\int_{0}^{\infty}db\;\frac{a}{2\pi^{3}}\frac{16+(a^{4}+1+b^{2})(a^{4}-13+b^{2})}{(a^{4}+1+b^{2})^{3}} \nonumber
\frac{\coth\left(	\sqrt{(C_{\perp_{2}}^{2}a^{2}+C_{z_{2}}^{2})/b}	\right)}{\sqrt{C_{\perp_{1}}^{2}a^{2}+C_{z_{1}}^{2}}}\\
=&\frac{C_{t_{z}}}{N_{f}}\ln(\Lambda/\mu),
\end{align}
where $C_{t_{z}}=0.341231$. Note that $\delta_{t_{z}}$ has a logarithmic divergence both in the UV and IR cutoffs.

\subsection{Vertex correction}
The correction $\delta_{g}$ is given by
\begin{align}
\delta_{g}=&(ig)^{2}\int_{\Omega,\bm{q}}\frac{1}{2}\text{Tr}[G_{0}(i\omega,\bm{q})G_{0}(i\omega,\bm{q})]D(i\omega,\bm{q}) \nonumber \\
=&-g^{2}\int_{\Omega,\bm{q}}\frac{-\Omega^{2}+t_{\perp}^{2}(q_{x}^{2}+q_{y}^{2})^{2}+t_{z}^{2}q_{z}^{4}}{\left[\Omega^{2}+t_{\perp}^{2}(q_{x}^{2}+q_{y}^{2})^{2}+t_{z}^{2}q_{z}^{4}\right]^{2}}\frac{\coth\left(\sqrt{(C_{\perp_{2}}^{2}(q_{x}^{2}+q_{y}^{2})+C_{z_{2}}^{2}\beta q_{z}^{2})t_{\perp}/\Omega}\right)}{\frac{N_{f}e^{2}}{\sqrt{t_{\perp}t_{z}}}\sqrt{C_{\perp_{1}}^{2}(q_{x}^{2}+q_{y}^{2})+C_{z_{1}}^{2}\beta q_{z}^{2}}}\nonumber \\
=&-\frac{\sqrt{t_{\perp}t_{z}}}{8\pi^{3}N_{f}t_{\perp}^{2}}\int_{-\infty}^{\infty}d\Omega \int_{\mu<|q_{z}|<\Lambda}dq_{z}\int_{-\infty}^{\infty}dq_{\perp}\;q_{\perp}\frac{-\Omega^{2}/t_{\perp}^{2}+q_{\perp}^{4}+\beta^{2}q_{z}^{4}}{(\Omega^{2}/t_{\perp}^{2}+q_{\perp}^{4}+\beta^{2}q_{z}^{4})^{2}}\frac{\coth\left(\sqrt{(C_{\perp_{2}}^{2}q_{\perp}^{2}+C_{z_{2}}^{2}\beta q_{z}^{2})t_{\perp}/\Omega}\right)}{\sqrt{C_{\perp_{1}}^{2}q_{\perp}^{2}+C_{z_{1}}^{2}\beta q_{z}^{2}}}.
\end{align}
After changing the integration variables, $q_{\perp}\rightarrow \sqrt{\beta}q_{z}a$ and $\Omega\rightarrow \beta t_{\perp} q_{z}^{2}b$, we have
\begin{align}
\delta_g=&\frac{(t_{\perp}t_{z})^{1/2}}{N_{f}}\frac{1}{t_{\perp}^{2}}\frac{t_{\perp}\beta^{2}}{\beta^{5/2}}\ln(\Lambda/\mu)\int_{0}^{\infty}da\int_{0}^{\infty}db\;\frac{a}{2\pi^{3}}\frac{-a^{2}-1+b^{2}}{(a^{4}+1+b^{2})^{2}}
\frac{\coth\left(	\sqrt{(C_{\perp_{2}}^{2}a^{2}+C_{z_{2}}^{2})/b}	\right)}{\sqrt{C_{\perp_{1}}^{2}a^{2}+C_{z_{1}}^{2}}}\nonumber\\
=&\frac{C_{g}}{N_{f}}\ln(\Lambda/\mu),
\end{align}
where $C_{g}=C_{\omega}$, which is consistent with the Ward identity.

\section{Consistency between the large $N_{f}$ calculation and $\epsilon$ expansion}
In this section, we will show the correspondence between the large $N_{f}$ calculation and the $\epsilon$ expansion.\\
In the static ($\Omega=0$) and long wavelength limit ($q\rightarrow0$), the boson propagator in the large $N_{f}$ approximation has the following form for the momentum dependence:
\begin{align}
D(i\omega=0,\bm{q}\rightarrow0)^{-1}\sim q_{\perp}+|q_{z}|.
\end{align}

Let us consider the $\epsilon$ expansion case. In the $\epsilon$ expansion, near the interacting fixed point,
\begin{align}
\alpha^{*}\gamma^{*}=&\frac{\epsilon}{N_{f}}\left(1-\frac{c_{N_{f}}}{N_{f}}\right)\approx\frac{\epsilon}{N_{f}},\\
\frac{\alpha^{*}}{\gamma^{*}}=&\frac{\epsilon}{N_{f}}\frac{1}{1-c_{N_{f}}/N_{f}}\approx\frac{\epsilon}{N_{f}}\left(1+\frac{c_{N_{f}}}{N_{f}}\right)\approx\frac{\epsilon}{N_{f}},
\end{align}
where we only keep up to $N_{f}^{-1}$ order because we consider the large $N_{f}$ limit. Using these results, 
\begin{align}
D(i\omega=0,\bm{q}\rightarrow0)^{-1}=&aq_{\perp}^{2}+\frac{1}{a}q_{z}^{2}-\Pi(i\omega,\bm{q})\nonumber\\
=&a\left(1+N_{f}\frac{\alpha^{*}}{\gamma^{*}}\ell\right)q_{\perp}^{2}+\frac{1}{a}\left(1+N_{f}\alpha^{*}\gamma^{*}\ell\right)q_{z}^{2}\nonumber\\
\approx&a\left(1+\epsilon\ell\right)q_{\perp}^{2}+\frac{1}{a}\left(1+\epsilon\ell\right)q_{z}^{2}\nonumber\\
\sim&e^{\epsilon\ell}q_{\perp}^{2}+e^{\epsilon\ell}q_{z}^{2}\nonumber\\
\approx&q_{\perp}^{2-\epsilon/z_{\perp}}+|q_{z}|^{2-\epsilon}.
\end{align}
Here, in the fourth line, we absorbed the momentum dependence of $a$ into $q_{\perp}$ and $q_z$.
For a sufficiently large $N_{f}$, $z_{\perp}\approx1$, thus for $\epsilon=1$ with $d=3$, $D(0,\bm{q})^{-1}\sim q_{\perp}+|q_{z}|$. Therefore, the result of the $\epsilon$ expansion is consistent with the large $N_{f}$ calculation.

\section{Physical observables in the non-interacting limit}\label{app:observable}

In this section, we will calculate the physical observables such as the specific heat, compressibility, diamagnetic susceptibility, and optical conductivity at the TQPT between DWSM and insulating phases in the non-interacting limit. For simplicity, we assume $t_{x}=t_{y}=t_{\perp}$, the rotational symmetry along the $k_z$-axis.

\subsection{Density of states}
Through the analytic continuation $i\omega \rightarrow \omega+i\delta$ in $G_{0}(i\omega,\bm{k})$, the retarded Green's function $G_{0}^{\text{ret}}$ is obtained as
\begin{align}
G^{\text{ret}}_{0}(\omega+i\delta,\bm{k})=&\frac{1}{-(\omega+ i\delta)+\mathcal{H}_{0}(\bm{k})},
\end{align}
and the imaginary part of $G_{0}^{\text{ret}}$ and the spectral function are
\begin{align}
\text{Im}[G^{\text{ret}}_{0}(\omega,\bm{k})]=&\frac{\pi\text{sgn}(\omega)}{2E_{k}}(\omega+\mathcal{H}_{0}(\bm{k}))\left(\delta(\omega-E_{k})+\delta(\omega+E_{k})\right),\\
S_{F}(\omega)=&-\frac{1}{\pi}\text{Tr}[G^{\text{ret}}_{0}(\omega,\bm{k})]\nonumber\\
=&\delta(\omega+E_{k})+\delta(\omega-E_{k}).
\end{align}
The density of states is given by
\begin{align}
\rho(\omega)=&\int\frac{d^{3}k}{(2\pi)^{3}}S_{F}(\omega,\bm{k})\nonumber\\
=&\frac{|\omega|}{\pi^{2}}\int_{0}^{\infty} dk_{\perp}\int_{0}^{\infty}dk_{z}\;k_{\perp}\delta(\omega^{2}-(t_{\perp}^{2}k_{\perp}^{4}+t_{z}^{2}k_{z}^{4}))\nonumber\\
=&\frac{\Gamma(5/4)}{4\pi^{3/2}\Gamma(3/4)}\frac{|\omega|^{1/2}}{t_{\perp}t_{z}^{1/2}},
\end{align}
where $\Gamma(x)$ is the gamma function and we use the identity,
\begin{align}
\int_{0}^{1} dR\;\frac{R}{(1-R^{4})^{3/4}}=&\frac{\sqrt{\pi}\Gamma(5/4)}{\Gamma(3/4)}.
\end{align}

\subsection{Free energy}
In this section, we will calculate the free energy at the TQPT in the non-interacting limit from which the specific heat and the compressibility are derived. The finite-temperature propagator of fermion is
\begin{align}
G_{0}(i\omega_{n},\bm{k})^{-1}=&(-i\omega_{n}-\mu)+\mathcal{H}_{0}(\bm{k}),
\end{align}
where we introduce the chemical potential $\mu$ for deriving the compressibility. The partition function and its logarithmic form are given by
\begin{align}
\mathcal{Z}=&\text{Det}[\beta G_{0}^{-1}]\nonumber\\
=&\prod_{i\omega_{n}}\prod_{\bm{k}}\left[\beta^{2}((\omega_{n}-i\mu)^{2}+E(\bm{k})^{2})\right],\\
\ln\mathcal{Z}=&V\int\frac{d^{3}k}{(2\pi)^{3}}T\sum_{i\omega_{n}}\ln\left[\beta^{2}((\omega_{n}-i\mu)^{2}+E(\bm{k})^{2})\right]\nonumber\\
=&\frac{V}{2}\int\frac{d^{3}k}{(2\pi)^{3}}T\sum_{i\omega_{n}}\left[\ln\left\{\beta^{2}(\omega_{n}^{2}+(E(\bm{k})-\mu)^{2})\right\}+\ln\left\{\beta^{2}(\omega_{n}^{2}+(E(\bm{k})+\mu)^{2})\right\}\right],
\end{align}
where $\beta=T^{-1}$ and we use the relation
\begin{align}
\left[(\omega_{n}-i\mu)^{2}+E(\bm{k})^{2}\right]\left[(\omega_{n}+i\mu)^{2}+E(\bm{k})^{2}\right]=&\left[\omega_{n}^{2}+(E(\bm{k})-\mu)^{2}\right]\left[\omega_{n}^{2}+(E(\bm{k})+\mu)^{2}\right].
\end{align}
By using
\begin{align}
\sum_{i\omega_{n}}\ln\left[\beta^{2}(\omega_{n}^{2}+E(\bm{k})^{2})\right]=& E(\bm{k})/T+2\ln(1+e^{-E(\bm{k})/T})+\text{const.},
\end{align}
we obtain the free energy density as
\begin{align}
\mathcal{F}=&-\frac{T}{V}\ln\mathcal{Z}\nonumber\\
=&-T\int\frac{d^{3}k}{(2\pi)^{3}}\left[	E(\bm{k})/T+\ln(1+e^{-(E(\bm{k})-\mu)/T})+\ln(1+e^{-(E(\bm{k})+\mu)/T})+\text{const.}	\right].
\end{align}
Subtracting $T=0$ contribution, $\delta\mathcal{F}(T):=\mathcal{F}(T)-\mathcal{F}(0)$ is given by
\begin{align}
\delta\mathcal{F}(T,\mu)=&-T\int\frac{d^{3}k}{(2\pi)^{3}}\left[	\ln(1+e^{-(E(\bm{k})-\mu)/T})+\ln(1+e^{-(E(\bm{k})+\mu)/T})	\right]\nonumber\\
=&\frac{\Gamma(5/4)}{8\pi\Gamma(3/4)}\frac{T^{5/2}}{t_{\perp}t_{z}^{1/2}}\left[\text{Li}_{\frac{5}{2}}(-e^{\mu/T})+\text{Li}_{\frac{5}{2}}(-e^{-\mu/T})\right],
\end{align}
where $\text{Li}_{n}(x)$ is the polylogarithm function.

\subsubsection{Specific heat}
For $\mu=0$, using $\text{Li}_{\frac{5}{2}}(-1)=-\frac{(4-\sqrt{2})}{4}\zeta(5/2)$ with the zeta function $\zeta(x)$, we get the free energy $\delta\mathcal{F}(T,0)$ as
\begin{align}
\delta\mathcal{F}(T,0)
=&-\frac{(4-\sqrt{2})\Gamma(5/4)\zeta(5/2)}{16\pi\Gamma(3/4)}\frac{T^{5/2}}{t_{\perp}t_{z}^{1/2}}.
\end{align}
The specific heat at $\mu=0$ is then given by
\begin{align}
C_{V}=&-T\frac{\partial^{2}\delta\mathcal{F}(T,0)}{\partial T^{2}}\nonumber\\
=&\frac{15(4-\sqrt{2})\Gamma(5/4)\zeta(5/2)}{64\pi\Gamma(3/4)}\frac{T^{3/2}}{t_{\perp}t_{z}^{1/2}}.
\end{align}

\subsubsection{Compressibility}
The compressibility is given by
\begin{align}
\kappa=&-\frac{\partial^{2}\delta\mathcal{F}(T,\mu)}{\partial\mu^{2}}\nonumber\\
=&-\frac{\Gamma(5/4)}{8\pi\Gamma(3/4)}\frac{T^{1/2}}{t_{\perp}t_{z}^{1/2}}\left[\text{Li}_{\frac{1}{2}}(-e^{\mu/T})+\text{Li}_{\frac{1}{2}}(-e^{-\mu/T})\right].
\end{align}
At $\mu=0$, we have
\begin{align}
\kappa
=&-\frac{(\sqrt{2}-1)\Gamma(5/4)\zeta(1/2)}{4\pi\Gamma(3/4)}\frac{T^{1/2}}{t_{\perp}t_{z}^{1/2}},
\end{align}
where $\text{Li}_{1/2}(-1)=(\sqrt{2}-1)\zeta(1/2)$ is used. Note that $\zeta(1/2)<0$, hence, $\kappa >0$.
\begin{align}
\end{align}

\subsection{Diamagnetic susceptibility}
Using the Fukuyama formula \cite{smFUKUYAMA1970111}, the diamagnetic susceptibility is given by
\begin{align}
\chi_{D,x}=&e_{0}^{2}T\sum_{i\omega_{n}}\int\frac{d^{3}k}{(2\pi)^{3}}\text{Tr}[J_{j}G(i\omega_{n},\bm{k})J_{k}G(i\omega_{n},\bm{k})J_{j}G(i\omega_{n},\bm{k})J_{k}G(i\omega_{n},\bm{k})],
\end{align}
where $J_{i}\equiv\frac{\partial\mathcal{H}_{0}}{\partial k_{i}}$ is the current operator,
\begin{align}
J_{x}=&2t_{\perp}k_{x}\sigma_{x}+2t_{\perp}k_{y}\sigma_{y},\\
J_{y}=&-2t_{\perp}k_{y}\sigma_{x}+2t_{\perp}k_{x}\sigma_{y},\\
J_{z}=&2t_{z}k_{z}\sigma_{z}.
\end{align}
Note that because of the $C_4$ symmetry of the Hamiltonian, $\chi_{D,x}=\chi_{D,y}=\chi_{D,\perp}$. Subtracting the zero temperature contribution to obtain a finite result, we have
\begin{align}
\chi_{D,\perp}=&e_{0}^{2}T\sum_{i\omega_{n}}\int\frac{d^{3}k}{(2\pi)^{3}}\text{Tr}[J_{y}G(i\omega_{n},\bm{k})J_{z}G(i\omega_{n},\bm{k})J_{y}G(i\omega_{n},\bm{k})J_{z}G(i\omega_{n},\bm{k})]\nonumber\\
&-e_{0}^{2}\int\frac{d\omega d^{3}k}{(2\pi)^{4}}\text{Tr}[J_{y}G(i\omega,\bm{k})J_{z}G(i\omega,\bm{k})J_{y}G(i\omega,\bm{k})J_{z}G(i\omega,\bm{k})],\nonumber\\
=&e_{0}^{2}t_{\perp}^{2}t_{z}^{2}\int\frac{d^{3}k}{(2\pi)^{3}}\left[	-32(k_{x}^{2}+k_{y}^{2})k_{z}^{2}M_{2}+128t_{\perp}^{2}t_{z}^{2}(k_{x}^{2}+k_{y}^{2})^{3}k_{z}^{6}M_{4}	\right]\nonumber\\
=&e_{0}^{2}t_{z}^{1/2}T^{1/2}c_{\chi,\perp},
\end{align}
where $c_{\chi,\perp}=0.054$. Here, we use
\begin{align}
\int_{0}^{\pi/2}\cos\theta_{R}\sin^{1/2}\theta_{R}\;d\theta_{R}=&\frac{2}{3},\\
\int_{0}^{\pi/2}\cos^{3}\theta_{R}\sin^{5/2}\theta_{R}\;d\theta_{R}=&\frac{8}{77},
\end{align}
and the following Matsubara frequency summations (where the zero-temperature contribution has been subtracted)
\begin{align}
M_{1}(\xi/T)=&T\sum_{i\omega_{n}}\frac{1}{(\omega_{n}^{2}+\xi^{2})}-\int_{-\infty}^{\infty}\frac{d\omega}{2\pi}\frac{1}{(\omega^{2}+\xi^{2})}\nonumber\\
=&\frac{1}{2\xi}\left[\tanh\left(\frac{\xi}{2T}\right)-1\right],\\
M_{2}(\xi/T)=&T\sum_{i\omega_{n}}\frac{1}{(\omega_{n}^{2}+\xi^{2})^{2}}-\int_{-\infty}^{\infty}\frac{d\omega}{2\pi}\frac{1}{(\omega^{2}+\xi^{2})^{2}}\nonumber\\
=&\frac{1}{4\xi^{3}}\left[\tanh\left(\frac{\xi}{2T}\right)-1\right]-\frac{1}{8\xi^{2}T}\frac{1}{\cosh^{2}(\frac{\xi}{2T})},\\
M_{3}(\xi/T)=&T\sum_{i\omega_{n}}\frac{1}{(\omega_{n}^{2}+\xi^{2})^{3}}-\int_{-\infty}^{\infty}\frac{d\omega}{2\pi}\frac{1}{(\omega^{2}+\xi^{2})^{3}}\nonumber\\
=&\frac{3}{16\xi^{5}}\left[\tanh\left(\frac{\xi}{2T}\right)-1\right]-\frac{3}{32\xi^{4}T}\frac{1}{\cosh^{2}(\frac{\xi}{2T})}-\frac{1}{32\xi^{3}T^{2}}\frac{\tanh\left(\frac{\xi}{2T}\right)}{\cosh^{2}\left(\frac{\xi}{2T}\right)},\\
M_{4}(\xi/T)=&T\sum_{i\omega_{n}}\frac{1}{(\omega_{n}^{2}+\xi^{2})^{4}}-\int_{-\infty}^{\infty}\frac{d\omega}{2\pi}\frac{1}{(\omega^{2}+\xi^{2})^{4}}\nonumber\\
=&\frac{5}{32\xi^{7}}\left[\tanh\left(\frac{\xi}{2T}\right)-1\right]-\frac{5}{64\xi^{6}T}\frac{1}{\cosh^{2}\left(\frac{\xi}{2T}\right)}-\frac{1}{32\xi^{5}T^{2}}\frac{\tanh\left(\frac{\xi}{2T}\right)}{\cosh^{2}\left(\frac{\xi}{2T}\right)}\nonumber\\
&+\frac{1}{384\xi^{4}T^{3}}\frac{1}{\cosh^{4}\left(\frac{\xi}{2T}\right)}\left[2-\cosh\left(\frac{\xi}{T}\right)\right].
\end{align}
Similarly, $\chi_{D,z}$ is given by
\begin{align}
\chi_{D,z}=&e_{0}^{2}T\sum_{i\omega_{n}}\int\frac{d^{3}k}{(2\pi)^{3}}\text{Tr}[J_{x}G(i\omega_{n},\bm{k})J_{z}G(i\omega_{n},\bm{k})J_{x}G(i\omega_{n},\bm{k})J_{z}G(i\omega_{n},\bm{k})]\nonumber\\
&-e_{0}^{2}\int\frac{d\omega d^{3}k}{(2\pi)^{4}}\text{Tr}[J_{x}G(i\omega,\bm{k})J_{z}G(i\omega,\bm{k})J_{x}G(i\omega,\bm{k})J_{z}G(i\omega,\bm{k})]\nonumber\\
=&e_{0}^{2}t_{\perp}^{4}\int\frac{d^{3}k}{(2\pi)^{3}}\left[	-32(k_{x}^{2}+k_{y}^{2})^{2}M_{2}+256t_{\perp}^{4}(k_{x}^{2}+k_{y}^{2})^{4}k_{x}^{2}k_{y}^{2}M_{4}	\right],\nonumber\\
=&\frac{e_{0}^{2}t_{\perp}}{t_{z}^{1/2}}T^{1/2}c_{\chi,z},
\end{align}
where $c_{\chi,z}=0.107$. Here, we used
\begin{align}
\int_{0}^{\pi/2}d\theta\;\frac{\cos^{2}\theta}{\sqrt{\sin\theta}}=&\frac{4\pi^{1/2}\Gamma(5/4)}{3\Gamma(3/4)},\\
\int_{0}^{\pi/2}d\theta\;\frac{\cos^{6}\theta}{\sqrt{\sin\theta}}=&\frac{80\pi^{1/2}\Gamma(5/4)}{77\Gamma(3/4)}.
\end{align}
In summary,
\begin{eqnarray}
\chi_{D,\perp}=&c_{\chi,\perp}e_{0}^{2}t_{z}^{1/2}T^{1/2},\quad
\chi_{D,z}=&c_{\chi,z}\frac{e_{0}^{2}t_{\perp}}{t_{z}^{1/2}}T^{1/2}.
\end{eqnarray}

\subsection{Optical conductivity}
The optical conductivity is given by
\begin{align}
\sigma_{ij}(\Omega,T)
=&e_{0}^{2}\int_{-\infty}^{\infty}\frac{d\omega}{\pi}\frac{n_{F}(\omega)-n_{F}(\omega+\Omega)}{\Omega}\int\frac{d^{3}k}{(2\pi)^{3}}\text{Tr}\left[J_{i}\text{Im}[G_{0}^{\text{ret}}(\omega,\bm{k})]J_{j}\text{Im}[G_{0}(\omega+\Omega,\bm{k})]\right],
\end{align}
where $n_{F}(x)=\frac{1}{1+e^{x/T}}$. Because of the $C_4$ symmetry of the Hamiltonian, $\sigma_{xx}=\sigma_{yy}$. Hence, we only need to consider $\sigma_{xx}$ and $\sigma_{zz}$.
\begin{align}
\sigma_{xx}(\Omega,T)
=&e_{0}^{2}\int_{-\infty}^{\infty}\frac{d\omega}{\pi}\;\frac{n_{F}(\omega)-n_{F}(\omega+\Omega)}{\Omega}\int\frac{d^{3}k}{(2\pi)^{3}}\text{Tr}\left[J_{x}\text{Im}[G_{0}^{\text{ret}}(\omega,\bm{k})]J_{x}\text{Im}[G_{0}^{\text{ret}}(\omega+\Omega,\bm{k})]\right]\nonumber\\
=&\frac{e_{0}^{2}T^{3/2}}{5t_{z}^{1/2}}\delta(\Omega)\int_{0}^{\infty}dR\;\frac{R^{3/2}}{\cosh^{2}\left(\frac{R}{2}\right)}+\frac{3}{20\sqrt{2}\pi}\frac{e_{0}^{2}}{t_{z}^{1/2}}|\Omega|^{1/2}\tanh\left(\frac{|\Omega|}{4T}\right),\\
\sigma_{zz}(\Omega,T)
=&e_{0}^{2}\int_{-\infty}^{\infty}\frac{d\omega}{\pi}\;\frac{n_{F}(\omega)-n_{F}(\omega+\Omega)}{\Omega}\int\frac{d^{3}k}{(2\pi)^{3}}\text{Tr}\left[J_{z}\text{Im}[G_{0}^{\text{ret}}(\omega,\bm{k})]J_{z}\text{Im}[G_{0}(\omega+\Omega,\bm{k})]\right]\nonumber\\
=&\frac{e_{0}^{2}T^{3/2}}{t_{\perp}t_{z}^{-1/2}}\frac{3\Gamma(-1/4)^{2}}{160\sqrt{2}\pi^{5/2}}\delta(\Omega)\int_{0}^{\infty}dR\;\frac{R^{3/2}}{\cosh^{2}\left(\frac{R}{2}\right)}+\frac{\sqrt{\pi}\Gamma(3/4)}{40\sqrt{2}\Gamma(5/4)}\frac{e_{0}^{2}}{t_{\perp}t_{z}^{-1/2}}|\Omega|^{1/2}\tanh\left(\frac{|\Omega|}{4T}\right).
\end{align}
Here, we used the following identities,
\begin{align}
\int_{0}^{\infty}dR\;\frac{R^{3/2}}{\cosh^{2}\left(\frac{R}{2}\right)}=&4.06856,\\
\int_{0}^{\pi/2}\frac{\cos^{7/2}\theta}{\sqrt{\cos\theta\sin\theta}}d\theta=&\frac{8}{5},\\
\int_{0}^{\pi/2}\sin^{5/2}\theta\;d\theta=&\frac{3\Gamma(-1/4)^{2}}{40\sqrt{2\pi}},\\
\int_{0}^{1}dR\;\frac{R^{3}(R^{4}-2)}{(1-R^{4})^{3/4}}=&-\frac{6}{5},\\
\int_{0}^{1}dR\;\frac{R^{5}}{\sqrt{1-R^{4}}}=&\frac{\sqrt{\pi}\Gamma(3/4)}{10\Gamma(5/4)},\\
\lim_{\Omega\rightarrow0}\frac{n_{F}(A)-n_{F}(A\pm\Omega)}{\Omega}=&\pm\frac{1}{4T}\frac{1}{\cosh^{2}(A/2T)}.
\end{align}
For $T=0$,
\begin{eqnarray}
\sigma_{xx}(\Omega)=&\frac{3}{20\sqrt{2}\pi}\frac{e_{0}^{2}}{t_{z}^{1/2}}|\Omega|^{1/2},\quad
\sigma_{zz}(\Omega)=&\frac{\sqrt{\pi}\Gamma(3/4)}{40\sqrt{2}\Gamma(5/4)}\frac{e_{0}^{2}}{t_{\perp}t_{z}^{-1/2}}|\Omega|^{1/2}.
\end{eqnarray}

\section{Effect of extra relevant perturbations}
In the presence of extra perturbations such as doping and disorder, a new parameter is introduced to characterize the extra perturbation in addition to the intrinsic length scale, correlation length $\xi$ set by temperature. For example, for doping, the Fermi wave vector $k_F$ is well defined. 
With the two parameters,  the two regimes naturally appear. For a large doping $k_F \xi\gg1$, our fixed point cannot be a good starting point, and it would be better to start from the Fermi liquid. On the other hand, $k_F \xi\ll1$, our description is certainly a good starting point and one can investigate the doping effect as a perturbation even though a little more additional cautions are necessary as in one of the standard critical phenomena.

\section{Sanity check of the power-law correction}\label{app:power}
In the main text, we included all the renormalization effects in the system parameters. Here, for a sanity check, equivalently we will include all the renormalization effects in the coordinates and obtain the associated anomalous dimensions. 

Recall that the RG equations for $t_{\perp}$ and $t_z$ are given by
\begin{align}
\frac{1}{t_{\perp}}\frac{dt_{\perp}}{d\ell}=&z-2z_{\perp}+\alpha F_{\perp}(\gamma),\\
\frac{1}{t_{z}}\frac{dt_{z}}{d\ell}=&z-2+\alpha F_{z}(\gamma).
\end{align}
Imposing $t_{\perp}$ and $t_{z}$ as constants, then we have
\begin{align}
z=&2-\alpha F_{z}(\gamma),\\
z_{\perp}=&1+\frac{\alpha}{2}\left[	F_{\perp}(\gamma)-F_{z}(\gamma)	\right].
\end{align}
At the fixed point $(\alpha,\gamma)=(\alpha^{*}$,$\gamma^{*})$, 
\begin{align}
z^{*}=&2-\alpha^{*} F_{z}(\gamma^{*}),\label{eq:zw}\\
z_{\perp}^{*}=&1+\frac{\alpha^{*}}{2}\left[	F_{\perp}(\gamma^{*})-F_{z}(\gamma^{*})	\right].\label{eq:zr}
\end{align}

Now, let us find the power-law corrections of the physical observables by using scaling hypothesis with the renormalized quantity $\mathcal{O}_{\text{R}}$ and the scaling dimension $d_\mathcal{O}$ for an observable $\mathcal{O}$. For the density of states, we have
\begin{align}
\rho=&b^{z-(2z_{\perp}+1)}\rho_{\text{R}},
\end{align}
whereas for the free energy,
\begin{align}
\mathcal{F}=&b^{-(z+2z_{\perp}+1)}\mathcal{F}_{\text{R}}.\label{eq:scalingfree-energy}
\end{align}
From Eq.~(\ref{eq:scalingfree-energy}), we obtain the specific heat and the compressibility, respectively, as
\begin{align}
C_{V}=&-T\frac{\partial^{2}\mathcal{F}}{\partial T^{2}}=b^{-(2z_{\perp}+1)}C_{V,\text{R}},\\
\kappa=&-\frac{\partial^{2}\mathcal{F}}{\partial \mu^{2}}=b^{z-(2z_{\perp}+1)}\kappa_{\text{R}}.
\end{align}

To determine the scaling relation of the optical conductivities and the diamagnetic susceptibilities, we use the minimal coupling $-i\partial_{i}\rightarrow -i\partial_{i}+e_{0}A_{i}(\tau,\bm{x})$, where $A_{i}(\tau,\bm{x})$ is a gauge-field. Since $e_{0}$ receives no renormalization at all, the scaling dimension of $A_{i}$ is the same as that of $\partial_{i}$. The optical conductivities and the diamagnetic susceptibilities can be obtained from the current-current response function $K_{ij}(i\omega,\bm{q})=\frac{1}{(2\pi)^{d+1}\delta(\omega+\Omega)\delta^{d}(\bm{q}+\bm{p})}\left\langle \mathcal{J}_{i}(i\omega,\bm{q})\mathcal{J}_{j}(i\Omega,\bm{p})\right\rangle$ with $\mathcal{J}_{i}(i\omega,\bm{q})=e_{0}\int_{\bm{k}}\psi^{\dagger}(i\omega,\bm{k+q})\frac{\partial \mathcal{H}_{0}(\bm{k})}{\partial k_{i}}\psi(i\omega,\bm{k})$ by the following relations \cite{smPhysRevLett.99.226803,smFUKUYAMA1970111}:
\begin{align}
\sigma_{ij}(\omega)=&\frac{1}{2\omega}\text{Im}K_{ij}^\text{ret}(\omega,\bm{q}=\bm{0}),\\
\chi_{D,i}(\omega)=&-\lim_{\bm{q}\rightarrow \bm{0}}\frac{\epsilon_{ijk}}{2q_{j}q_{k}}K_{jk}(0,\bm{q}).
\end{align}
Here, the repeated indices are not summed.
Because $\left\langle \mathcal{J}_{i}(i\omega,\bm{q})\mathcal{J}_{j}(i\Omega,\bm{p})\right\rangle$ is obtained by differentiating the logarithm of the partition function $Z[\bm{A}]$ with respect to $A_{i}(i\omega,\bm{q})$ and $A_{j}(i\omega,\bm{p})$, the scaling dimension of $K_{ij}(i\omega,\bm{q})$, namely $[K_{ij}]$, is given by
\begin{align}
[K_{ij}]=&[\frac{\delta}{\delta A_{i}(i\omega,\bm{q})}]+[\frac{\delta}{\delta A_{j}(i\omega,\bm{q})}]-[d\tau]-[d^{d}\bm{x}]\nonumber\\
=&-[\partial_{i}]-[\partial_{j}]+(z+2 z_{\perp}+d-2).
\end{align}
Equipped with this scaling relation of $K_{ij}$, we can derive the following relations:
\begin{align}
\sigma_{\perp\perp}&=b^{d-2}\sigma_{\perp\perp,\text{R}},\\
\sigma_{zz}&=b^{2z_{\perp}-d+2}\sigma_{zz,\text{R}},\\
\chi_{D,\perp}&=b^{-z+d-2}\chi_{D,\perp,\text{R}},\\
\chi_{D,z}&=b^{-z+2z_{\perp}-d+2}\chi_{D,z,\text{R}}.
\end{align} 
The RG equation of the temperature and frequency is
\begin{align}
\frac{d\mathcal{O}}{d\ln b}=z\mathcal{O},
\end{align}
where $\mathcal{O}=T,\Omega$. Let $z=z^{*}$ and $z_{\perp}=z_{\perp}^{*}$. Solving this, we obtain $\mathcal{O}(b)=b^{z^{*}}\mathcal{O}$. Let $b^{*}$ be the cutoff value, so that $\mathcal{O}(b^{*})=(b^{*})^{z^{*}}\mathcal{O}=\Lambda$, then $b^{*}=(\Lambda/\mathcal{O})^{1/z^{*}}\propto \mathcal{O}^{-1/z^{*}}$. Using this, we can obtain the power-law corrections of the observables in terms of the temperature and frequency.

For the density of states, we have
\begin{align}
\rho\propto |\Omega|^{(2z_{\perp}^{*}+1-z^{*})/z^{*}}\propto |\Omega|^{1/2+c_{\perp}+\frac{1}{2}c_{z}}.
\end{align}
For the specific heat and compressibility,
\begin{align}
C_{V}\propto T^{(2z_{\perp}^{*}+1)/z^{*}}\approx T^{3/2+c_{\perp}+\frac{1}{2}c_{z}},\\
\kappa\propto T^{(2z_{\perp}^{*}+1-z^{*})/z^{*}}\approx T^{1/2+c_{\perp}+\frac{1}{2}c_{z}}.
\end{align}
For the diamagnetic susceptibility,
\begin{align}
\chi_{D,\perp}\propto& T^{(z^{*}-1)/z^{*}}\approx T^{1/2-\frac{1}{2}c_{z}},\\
\chi_{D,z}\propto& T^{(z^{*}-2z_{\perp}^{*}+1)/z^{*}}\approx T^{1/2-c_{\perp}+\frac{1}{2}c_{z}}.
\end{align}
For the optical conductivity,
\begin{align}
\sigma_{xx}\propto& \Omega^{1/z^{*}}\approx \Omega^{1/2+c_{z}},\\
\sigma_{zz}\propto& \Omega^{(2z_{\perp}^{*}-1)/z^{*}}\approx \Omega^{1/2+c_{\perp}-\frac{1}{2}c_{z}}.
\end{align}
Here, $c_{\perp}\approx0.402/N_{f}$ and $c_z\approx0.044/N_{f}$ in the large $N_{f}$ approximation. Thus, we obtain the same results as in the main text. If the symmetry-allowed parabolic term is included, we have $c_{\perp}\approx0.145/N_{f}$ and $c_{z}\approx0.050/N_{f}$.

For the candidate materials of DWSM, HgCr$_{2}$Se$_{4}$ and SrSi$_{2}$, HgCr$_{2}$Se$_{4}$ has one pair ($N_{f}=1$) of double-Weyl points, whereas SrSi$_{2}$ has six pairs ($N_{f}=6$) of double-Weyl points. In particular, for SrSi$_{2}$, it has cubic symmetry, therefore, to see the anisotropic behaviors, we need to maintain only one $C_{4}$ symmetry. For example, if we apply a uniaxial pressure along $\hat{z}$, then the $C_{4}$ symmetry along $\hat{x}$ and $\hat{y}$ is broken, so we only have two pairs of double-Weyl points on the $\hat{z}$ axis \cite{smHuang1180}. 
Therefore, under this situation, the effective number of pairs of double-Weyl points of SrSi$_{2}$ is two ($N_{f}=2$). Then, for $\eta_{2}\equiv c_{z}/2$ and $\eta_{3}\equiv c_{\perp}-c_{z}/2$ we find that $\eta_2-\eta_3$ values for HgCr$_{2}$Se$_{4}$ and SrSi$_{2}$ are $-0.198$ and $-0.132$, respectively. We expect that the anisotropic scaling will be manifested at low temperatures or low frequencies.

\section{Stability of the anisotropic NFL fixed point under short-range interactions}\label{app:shortrange}
In this section, we study the effect of short-range interactions which was reported to destroy the non-Fermi liquid phase in the pyrochlore iridates $A_{2}\text{Ir}_{2}\text{O}_7$ \cite{smPhysRevLett.113.106401,smPhysRevB.92.045117} and show that the non-Fermi liquid phase of DWSM at the TQPT remains stable in a realizable range of $N_{f}$ in $d=3$.

\subsection{Possible type of short-range interactions}
Here, we investigate how short-range interactions affect the non-Fermi liquid we have found. We first note that 
\begin{equation}
\left(\psi^\dagger \sigma_{0} \psi\right)^{2}=-\left(\psi^\dagger \sigma_{i} \psi\right)^{2}=\frac{1}{2} \left(\psi^\dagger \sigma_{y} \psi^{*} \right) \left(\psi^{\intercal} \sigma_{y} \psi \right) \label{eq:identity}
\end{equation}
for $i=x,y,z$. Using this identity, we can study the effects of all possible short-range interactions in the particle-particle and particle-hole channels by adding just the following interaction to the action in Eq. (\ref{eq:action}):
\begin{align}
S_{u}=\frac{u}{2} \int{ {d}\tau {d}^d x} \left(\psi^\dagger \sigma_{0} \psi\right)^{2}_{(\tau,x)}.
\end{align}
In contrast to Ref.~\cite{smPhysRevLett.113.106401,smPhysRevB.92.045117} where the 4 by 4 gamma matrices are used and the vector-type short-range interactions appear, only the scalar-type interaction is needed in the present case.

\subsection{One-loop corrections}

\begin{figure}[t]
\centering
\subfigure[]{
\includegraphics[]{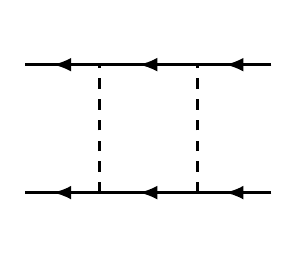}
\label{fig_ph:a}}
\subfigure[]{
\includegraphics[]{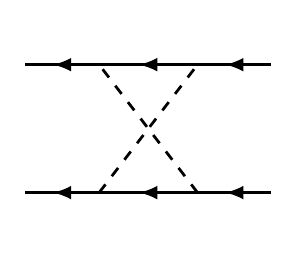}
\label{fig_ph:b}}
\subfigure[]{
\includegraphics[]{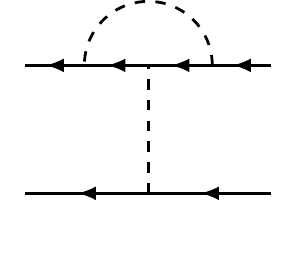}
\label{fig_ph:c}}
\subfigure[]{
\includegraphics[]{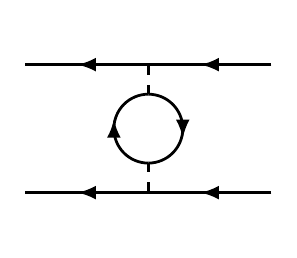}
\label{fig_ph:d}}\\
\subfigure[]{
\includegraphics[]{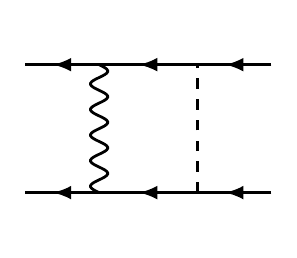}
\label{fig_ph:e}}
\subfigure[]{
\includegraphics[]{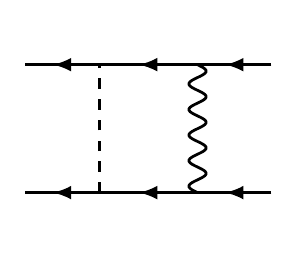}
\label{fig_ph:f}}
\subfigure[]{
\includegraphics[]{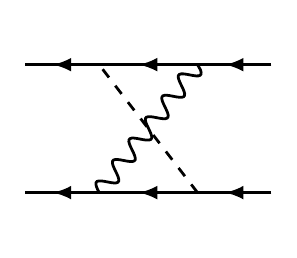}
\label{fig_ph:g}}
\subfigure[]{
\includegraphics[]{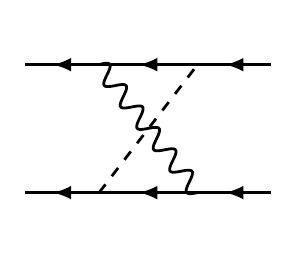}
\label{fig_ph:h}}
\subfigure[]{
\includegraphics[]{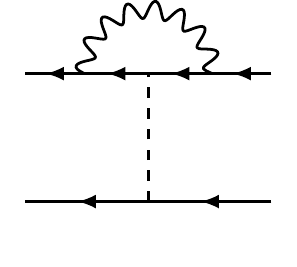}
\label{fig_ph:i}}
\subfigure[]{
\includegraphics[]{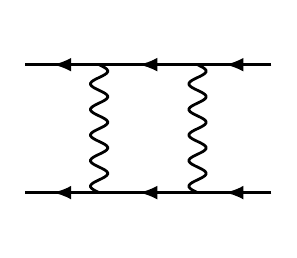}
\label{fig_ph:j}}
\subfigure[]{
\includegraphics[]{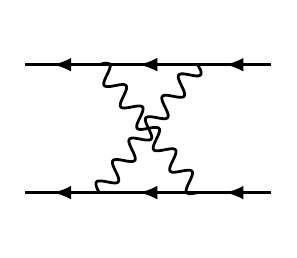}
\label{fig_ph:k}}
\caption{Feynman diagrams generated by the short-range interaction and long-range Coulomb interaction. The dashed and wavy lines stand for the short-range interaction and long-range Coulomb interaction, respectively. The solid line with an arrowhead stands for the fermion propagator.}\label{fig:shortrange_diagrams}
\end{figure}

To obtain the corrections generated by the short-range interaction and the combination of the short-range and Coulomb interactions up to one-loop order, we evaluate the Feynman diagrams in Fig.~\ref{fig:shortrange_diagrams}.
Among the diagrams in Fig.~\ref{fig:shortrange_diagrams}, only the diagrams Figs.~\ref{fig_ph:b}, \ref{fig_ph:g}, \ref{fig_ph:h}, and \ref{fig_ph:k} give us the following non-zero corrections to the short-range interaction, 
\begin{align*}
\delta u^{(b)} d\ell=&-u^{2}\int_{\partial\Lambda}\frac{d\Omega d^{d}q}{(2\pi)^{d+1}}\eta_{\mu\nu}\frac{\text{Tr}[\sigma_{0}G(i\Omega,\bm{q})\sigma_{0}\sigma_{\mu}]}{\text{Tr}[\sigma_{\mu}\sigma_{\mu}]}\frac{\text{Tr}[\sigma_{0}G(i\Omega,\bm{q})\sigma_{0}\sigma_{\mu}]}{\text{Tr}[\sigma_{\mu}\sigma_{\mu}]},\\
\delta u^{(g),(h)} d\ell=&u(ig)^{2}\int_{\partial\Lambda}\frac{d\Omega d^{d}q}{(2\pi)^{d+1}}\eta_{\mu\nu}\frac{\text{Tr}[\sigma_{0}G(i\Omega,\bm{q})\sigma_{0}\sigma_{\mu}]}{\text{Tr}[\sigma_{\mu}\sigma_{\mu}]}\frac{\text{Tr}[\sigma_{0}G(i\Omega,\bm{q})\sigma_{0}\sigma_{\mu}]}{\text{Tr}[\sigma_{\mu}\sigma_{\mu}]}D_{0}(i\Omega,\bm{q}),\\
\delta u^{(k)} d\ell=&-(ig)^{4}\int_{\partial\Lambda}\frac{d\Omega d^{d}q}{(2\pi)^{d+1}}\eta_{\mu\nu}\frac{\text{Tr}[\sigma_{0}G(i\Omega,\bm{q})\sigma_{0}\sigma_{\mu}]}{\text{Tr}[\sigma_{\mu}\sigma_{\mu}]}\frac{\text{Tr}[\sigma_{0}G(i\Omega,\bm{q})\sigma_{0}\sigma_{\mu}]}{\text{Tr}[\sigma_{\mu}\sigma_{\mu}]}D_{0}(i\Omega,\bm{q})^{2},\\
\end{align*}
where $\eta_{\mu\nu}\equiv\text{diag}(1,-1,-1,-1)$  and the repeated indices are summed over.  Here, by using Eq.~(\ref{eq:identity}), we convert the corrections to `vector-type' short-range interaction such as $(\psi^{\dagger}\sigma_{i}\psi)^{2}$ ($i=x,y,z$) into the corrections to the `scalar-type' short-range interaction such as $(\psi^{\dagger}\sigma_{0}\psi)^{2}$, which is reflected in $\eta_{\mu\nu}$.
As a result, we obtain the following correction $\delta u$ to $u$:
\begin{align}
\delta u d\ell=&\delta u^{(b)}d\ell+\delta u^{(g)}d\ell+\delta u^{(h)}d\ell+\delta u^{(k)}d\ell \nonumber\\
=&-\sum_{a,b=u,g}\int_{\partial\Lambda}\frac{d\Omega d^{d}q}{(2\pi)^{d+1}}\eta_{\mu\nu}\frac{\text{Tr}[\sigma_{0}G(i\Omega,\bm{q})\sigma_{0}\sigma_{\mu}]}{\text{Tr}[\sigma_{\mu}\sigma_{\mu}]}\frac{\text{Tr}[\sigma_{0}G(i\Omega,\bm{q})\sigma_{0}\sigma_{\mu}]}{\text{Tr}[\sigma_{\mu}\sigma_{\mu}]}I_{a}(\bm{q})I_{b}(\bm{q}) \nonumber \\
=&\frac{u^{2}}{4}\int_{\partial \Lambda}\frac{{d}^{d}q}{\left(2\pi\right)^{d}}\frac{1}{\varepsilon_{\bm{q}}} + \frac{u g^2}{2}\int_{\partial \Lambda}\frac{{d}^{d}q}{\left(2\pi\right)^{d}}\frac{1}{a q_\perp^2+\frac{1}{a}q_z^2}\frac{1}{\varepsilon_{\bm{q}}} +\frac{g^4}{4} \int_{\partial \Lambda}\frac{{d}^{d}q}{\left(2\pi\right)^{d}}\frac{1}{(a q_\perp^2+\frac{1}{a}q_z^2)^2}\frac{1}{\varepsilon_{\bm{q}}}, \label{eq:deltag0_tentative}
\end{align}
where $I_{u}(\bm{q})=u$ and $I_{g}(\bm{q})=\frac{g^{2}}{a(k_{x}^{2}+k_{y}^{2})+k_{z}^{2}/a}$. 
Introducing a dimensionless parameter
\begin{align}
\bar{u}=\frac{S_{d-2}}{4\pi\left(2\pi\right)^{d-2}\Lambda^{2-d}\left|t_{\perp}\right|}u, \nonumber 
\end{align}
the correction $\delta u$ to the dimensionless $\bar{u}$ is obtained from Eq. (\ref{eq:deltag0_tentative}),
\begin{align}
\delta u = \bar{u}^{2}H_{1}(\lambda)+\bar{u}\alpha H_{2}(\gamma,\lambda)+\alpha^{2}H_{3}(\gamma,\lambda),
\end{align} 
with 
\begin{align}
H_{1}(\gamma,\lambda)
=&\int_{0}^{\Lambda_{\rho}}\frac{\rho{d}\rho}{\sqrt{\rho^{4}+\left(1+\lambda\rho^{2}\right)^{2}}} \label{eq:G1}\\
=&\frac{1}{2\sqrt{1+\lambda^{2}}}\log\left[1+\frac{\sqrt{1+\left(\sqrt{1+\lambda^{2}}+\lambda\right)^{2}}}{\sqrt{1+\lambda^{2}}+\lambda}\right],\nonumber\\
H_{2}(\gamma,\lambda)
=&12\gamma\int_{0}^{\infty}\frac{\rho{d}\rho}{\left(1+4\gamma^{2}\rho^{2}\right)\sqrt{\rho^{4}+\left(1+\lambda\rho^{2}\right)^{2}}} \label{eq:G2}\\
=&\frac{6\gamma}{\sqrt{1+\left(4\gamma^{2}-\lambda\right)^{2}}}\log\left[\frac{4\gamma^{2}\left(4\gamma^{2}-\lambda+\sqrt{1+\left(4\gamma^{2}-\lambda\right)^{2}}\right)}{\sqrt{1+\lambda^{2}}\sqrt{1+\left(4\gamma^{2}-\lambda\right)^{2}}+\lambda\left(4\gamma^{2}-\lambda\right)-1}\right],\nonumber\\
H_{3}(\gamma,\lambda)
=&36\gamma^{2}\int_{0}^{\infty}\frac{\rho{d}\rho}{\left(1+4\gamma^{2}\rho^{2}\right)^{2}\sqrt{\rho^{4}+\left(1+\lambda\rho^{2}\right)^{2}}} \label{eq:G3}\\
=&18\gamma^{2}\left[\frac{4\gamma^{2}-\sqrt{1+\lambda^{2}}}{1+\left(4\gamma^{2}-\lambda\right)^{2}}+\frac{\left(4\gamma^{2}-\lambda\right)\lambda-1}{\left(1+\left(4\gamma^{2}-\lambda\right)^{2}\right)^{3/2}}\log\left[\frac{4\gamma^{2}\left\{ \sqrt{1+\left(4\gamma^{2}-\lambda\right)^{2}}-\left(4\gamma^{2}-\lambda\right)\right\} }{1-\lambda\left(4\gamma^{2}-\lambda\right)+\sqrt{\left(1+\lambda^{2}\right)\left(1+\left(4\gamma^{2}-\lambda\right)^{2}\right)}}\right]\right], \nonumber
\end{align}
where $\Lambda_\rho =\left(1+\lambda^{2}\right)^{-1/4}$ is introduced to regulate the UV divergence in Eq. (\ref{eq:G1}).

After rescaling the fields and space-time coordinates, we finally obtain the RG flow functions for $\alpha$, $\gamma$, $\lambda$ and $\bar{u}$:
\begin{align}
\frac{1}{\alpha}\frac{d\alpha}{d\ell}&=\epsilon-\frac{\alpha}{2}\left[N_{f}\left\{ \frac{1}{\gamma}\left(\frac{2+\lambda^{2}}{2}-\frac{\lambda(5+2\lambda^{2})}{4\sqrt{1+\lambda^{2}}}\right)+\gamma\left(\frac{1+2\lambda^{2}}{\sqrt{1+\lambda^{2}}}-2\lambda\right)\right\} +F_{+}(\gamma,\lambda)\right], \label{eq:shortrange_RGalpha}\\
\frac{1}{\gamma}\frac{d\gamma}{d\ell}&=\frac{\alpha}{2}\left[N_{f}\left\{ \frac{1}{\gamma}\left(\frac{2+\lambda^{2}}{2}-\frac{\lambda(5+2\lambda^{2})}{4\sqrt{1+\lambda^{2}}}\right)-\gamma\left(\frac{1+2\lambda^{2}}{\sqrt{1+\lambda^{2}}}-2\lambda\right)\right\} +F_{-}(\gamma,\lambda)\right],\label{eq:shortrange_RGgamma}\\
\frac{1}{\lambda}\frac{d\lambda}{d\ell}&=-\frac{\alpha}{\lambda}\bigg[4\gamma{}^{2}F_{z}(\gamma,\lambda)+\lambda F_{\perp}(\gamma,\lambda)\bigg],\label{eq:shortrange_RGlambda}\\
\frac{d\bar{u}}{d\ell}&=\left(\epsilon-2-\alpha F_{\perp}(\gamma,\lambda)\right)\bar{u}+\delta\bar{u} \nonumber \\
&=\left(\epsilon-2-\alpha F_{\perp}(\gamma,\lambda)+\alpha H_{2}(\gamma,\lambda)\right)\bar{u}+H_{1}(\gamma,\lambda)\bar{u}^{2}+H_{3}(\gamma,\lambda)\alpha^{2}. \label{eq:shortrange_RGg}
\end{align}
Note that no modification arises in Eqs. (\ref{eq:shortrange_RGalpha}), (\ref{eq:shortrange_RGgamma}), and (\ref{eq:shortrange_RGlambda}) even if we include $u$, because $u$ does not yield self-energy corrections to the fermion $\psi$ and boson $\phi$ at leading order. For the particle-particle channel, it has the same RG flow equations as the particle-hole channel because they have the same operator form.

\subsection{RG analysis of the short-range interaction}

\begin{figure}[t]
\centering
\subfigure[$\;N_{f}=4$]{
\includegraphics[scale=0.6]{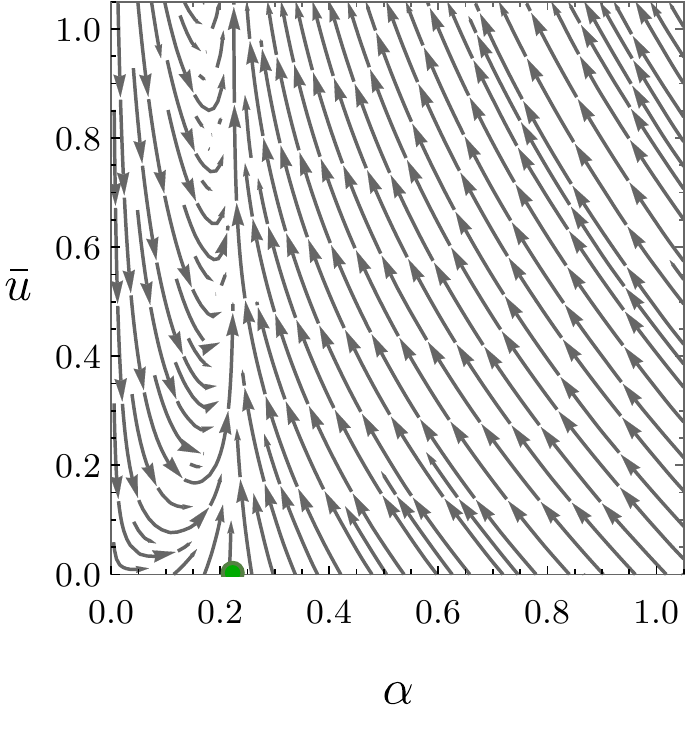}
}
\subfigure[$\;N_{f}=4.775$]{
\includegraphics[scale=0.6]{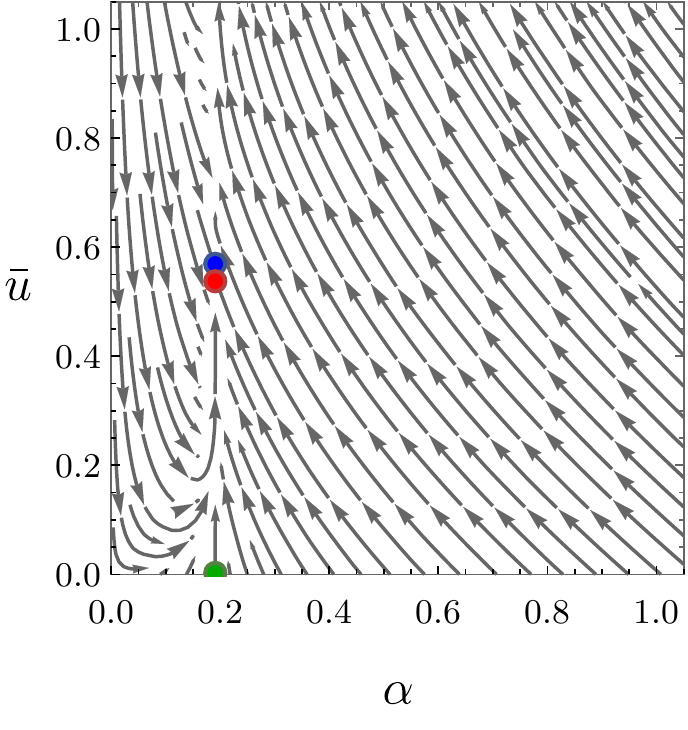}
}
\subfigure[$\;N_{f}=5$]{
\includegraphics[scale=0.6]{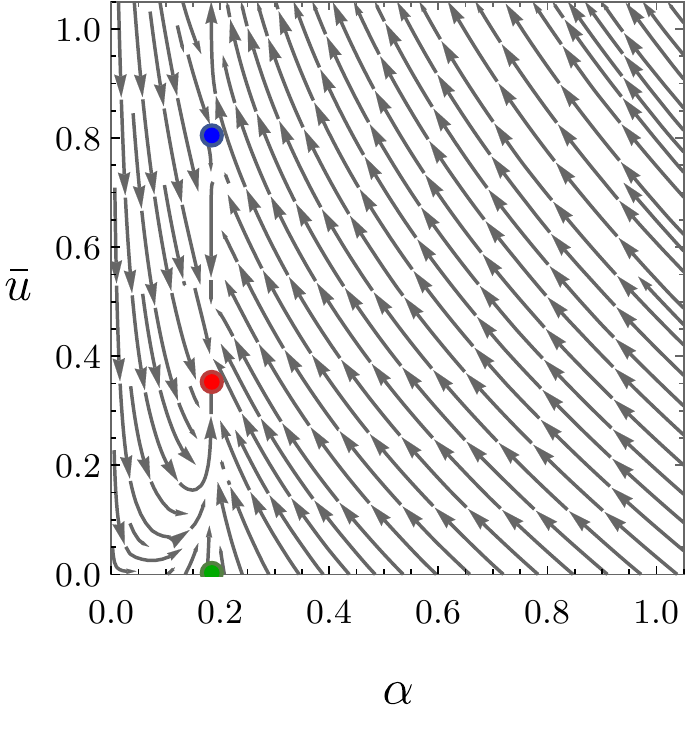}
}
\caption{RG flow diagrams in terms of $N_{f}$ when $s_\perp$ is ignored.}\label{fig:shortrange_RGflowdiagrams_nolambda}
\end{figure}

\begin{figure}[t]
\centering
\subfigure[$\;N_{f}=1$]{
\includegraphics[scale=0.6]{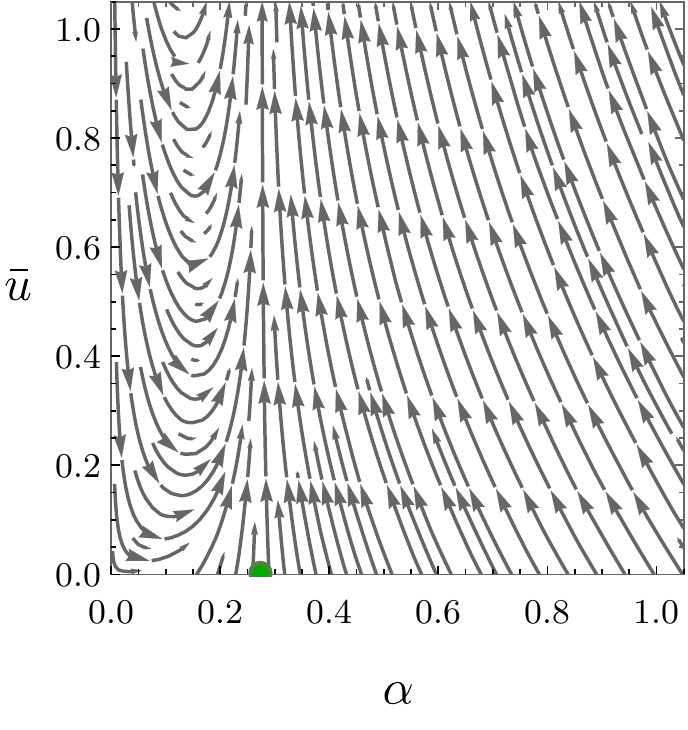}
}
\subfigure[$\;N_{f}=1.883$]{
\includegraphics[scale=0.6]{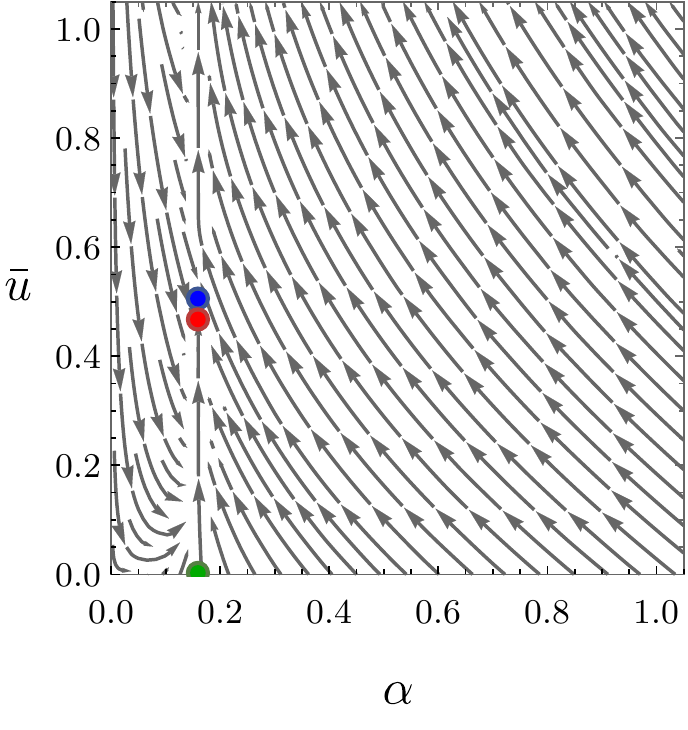}
}
\subfigure[$\;N_{f}=2$]{
\includegraphics[scale=0.6]{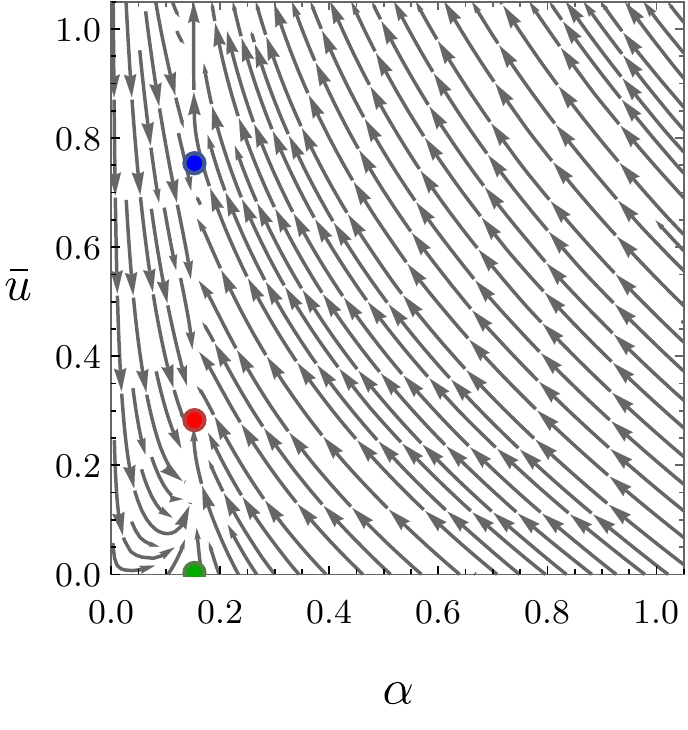}
}
\caption{RG flow diagrams in terms of $N_{f}$ when $s_\perp$ is included. }\label{fig:shortrange_RGflowdiagrams}
\end{figure}

The RG function for $\bar{u}$ in Eq.~(\ref{eq:shortrange_RGg}) includes a term proportional to $\alpha^2$. This $\alpha^2$ correction generates $\bar{u}$ during the RG even if we start with an initial condition $\bar{u}(\ell=0)=0$. Consequently, we find no stable fixed point with $N_{f}=1$ in $d=3$. First, let us ignore $\lambda$ in the RG functions of Eqs.~\cref{eq:shortrange_RGalpha,eq:shortrange_RGgamma,eq:shortrange_RGlambda,eq:shortrange_RGg}. Figure \ref{fig:shortrange_RGflowdiagrams_nolambda} shows the RG flow for $N_{f}=4$, $4.775$, and $5$ in $d=3$, where stable (unstable) fixed points are denoted by red (blue) points. The green point is the NFL fixed point obtained when the short-range interaction is neglected. As we can see, the lower bound of $N_{f}$, $N_{c}$, above which a stable fixed point begins to appear is approximately $N_{c}=4.775$. This result seems to show that the non-Fermi liquid phase of DWSM at the TQPT is not realizable since $N_{f}\ge5$ is not likely to be achievable in a real experiment. However, if we take $\lambda$ into account, we get a qualitatively different result. Figure \ref{fig:shortrange_RGflowdiagrams} shows the RG flow when $\lambda$ is kept. The estimated lower bound of $N_{f}$ is about $N_{c}=1.883$. Thus, a stable fixed point appear for $N_{f}=2$, which is much smaller compared to that obtained when $\lambda$ is neglected. We expect that DWSM at the TQPT with $N_{f}=2$ is accessible in an experiment with $\text{SrSi}_2$ \cite{smHuang1180}. Although the estimated values of the lower bound of $N_{f}$ are found to depend on the renormalization scheme, the symmetry-allowed $s_\perp$-term in DWSM at the TQPT seems to stabilize the non-Fermi liquid phase.

\subsection{Relevance of the short-range interaction in $\mathcal{O}(\epsilon)$ order}
In the previous section, we considered the corrections up to one-loop order. However, in the $\epsilon$ expansion, we find the anisotropic NFL fixed point up to order of $\mathcal{O}(\epsilon)$.  
For that reason, near our anisotropic NFL fixed point, $\alpha^{2}$ will give us the correction of $\mathcal{O}(\epsilon^{2})$. Therefore, if we carefully consider terms of order $\epsilon$, we can ignore $\alpha^{2}$ contributions in Eq.~(\ref{eq:shortrange_RGg}) near our anisotropic NFL fixed point. In this situation, the short-range interaction is irrelevant in $d=3$ for $N_f>N_{c}=2.279$ when $\lambda$ is neglected, whereas $N_{c}$ become $0.952$ if $\lambda$ is kept. Thus, keeping $\lambda$, we find that the non-Fermi liquid phase of DWSM at the TQPT with $N_f=1$ remains stable in the presence of the short-range interaction up to the accuracy of $\mathcal{O}(\epsilon)$. Note that to properly keep $\mathcal{O}(\epsilon^{2})$ contributions, we need to calculate the two-loop order diagrams, which is beyond the scope of the current work.

\end{document}